\input harvmac.tex

\input epsf.tex
\overfullrule=0mm
\newcount\figno
\figno=0
\def\fig#1#2#3{
\par\begingroup\parindent=0pt\leftskip=1cm\rightskip=1cm\parindent=0pt 
\baselineskip=11pt
\global\advance\figno by 1
\midinsert
\epsfxsize=#3
\centerline{\epsfbox{#2}}
{\bf Fig. \the\figno:} #1\par
\endinsert\endgroup\par
}
\def\figlabel#1{\xdef#1{\the\figno}}
\def\encadremath#1{\vbox{\hrule\hbox{\vrule\kern8pt\vbox{\kern8pt
\hbox{$\displaystyle #1$}\kern8pt}
\kern8pt\vrule}\hrule}}


\lref\Smir{F. Smirnov, ``Form factors in completely
integrable models of quantum field theory'', World Scientific and
references therein.}
\lref\Kar{M. Karowski, P. Weisz, Nucl. Phys. B139 (1978) 455.}
\lref\W{K. M. Watson, Phys. Rev. B95 (1954) 228.}
\lref\Korbook{V. E. Korepin, N.M Bogoliubov, A.G. Izergin, ``Quantum
inverse Scattering method and correlation functions", Cambridge
university press, 1993.}
\lref\Mussi{A. Fring, G. Mussardo and P. Simonetti, Phys. Lett. B307
(1993) 83; G. Mussardo, ``Spectral representation of correlation functions
in two-dimensional quantum field theories", Talk Int. Coll. QFT II, Tata insitute, hep-th/9405128.}
\lref\pkondo{P. Fendley, Phys. Rev. Lett. 71 (1993) 2485,
cond-mat/9304031.}
\lref\FSkon{P. Fendley and H. Saleur, Phys. Rev. Lett. 75 (1995) 4492.}
\lref\FLSnoise{P. Fendley, A. Ludwig and H. Saleur, Phys. Rev. Lett.
75 (1995) 2196.}
\lref\moon{K. Moon, H. Yi, C.L. Kane, S.M. Girvin and M.P.A. Fisher,
Phys.
Rev. Lett. 71 (1993) 4391, cond-mat/9408068.}
\lref\FSW{P. Fendley, H. Saleur, N. Warner, Nucl. Phys. B 430 (1994)
577,
hep-th/9406104.}
\lref\Affleck{I. Affleck, E. Sorensen, cond-mat/9508030
and cond-mat/9511031 to appear in Phys. Rev. B.}
\lref\thegoodguys{N. Andrei, K. Furuya and J. Lowenstein, Rev. Mod.
Phys. 55 (1983) 331; A.M. Tsvelick, P.B. Wiegmann, Adv. Phys. 32
(1983) 453.}
\lref\Schotte{ K.-D. Schotte, Z. Physik 230 (1970) 99.}
\lref\AYH{ P.W. Anderson, G. Yuval and D.R. Hamman, Phys. Rev. B1
(1970) 4464.}
\lref\Kondo{ J. Kondo, Prog. Th. Phys. 32 (1964) 37.}
\lref\FLSjack{ P. Fendley, F. Lesage and H. Saleur,
J. Stat. Phys. 79 (1995) 799, hep-th/9409176.}
\lref\Leggett{ A.J. Leggett, S. Chakravarty, A.T. Dorsey,
M.P.A. Fisher, A. Garg and W. Zwerger, Rev. Mod. Phys. 59
(1987) 1.}
\lref\chakra{S. Chakravarty and
J. Rudnick, Phys. Rev. Lett. 75 (1995) .}
\lref\KF{ C.L. Kane and M.P.A. Fisher, Phys. Rev. B46 (1992)
15233.}
\lref\FLS{P. Fendley, A.W.W. Ludwig and H. Saleur, Phys.
Rev. Lett. 74 (1995) 3005, cond-mat/9408068 and cond-mat/9503172.}
\lref\Schmid{ A. Schmid, Phys. Rev. Lett. 51 (1983) 1506.}
\lref\CFW{ C.de C. Chamon, D.E. Freed and X.G. Wen,
Phys. Rev. B51 (1995) 2363, cond-mat/9408064.}
\lref\Natan{N. Andrei, Phys. Lett.}
\lref\lem{K. Leung, R. Egger and C. Mak., Phys. Rev. B (1996).}
\lref\GZ{S. Ghoshal, A. Zamolodchikov, Int. J. Phys. A {\bf 9},
3841 (1994).}

\Title{USC-96-06}
{\vbox{
\centerline{Form factors approach to current correlations in}
 \vskip 4pt
\centerline{one dimensional systems with impurities.}}}

\centerline{F. Lesage, H. Saleur\footnote{$^\dagger$}
{Packard Fellow} and S. Skorik}
\bigskip\centerline{Department of Physics}
\centerline{University of Southern California}
\centerline{Los Angeles, CA 90089-0484}

\vskip .3in

We  show how to  compute analytically  time and space
dependent correlations in one
dimensional  quantum integrable systems with an impurity. Our
approach
is based  on a description of these systems in terms of massless
scattering of quasiparticles. Correlators follow then
from matrix
elements of local operators between multiparticle states, the
  ``massless form factors''. Although an infinite sum of
these form factors
has to be considered in principle, we find that for
current, spin, and energy operators, only a few
(typically two or three) are necessary to obtain an accuracy of more
than $1\%$, for {\bf arbitrary coupling strength}, that is all the
way from short to large distances. As examples we compute, at zero
temperature, the
frequency dependent conductance in a Luttinger liquid with impurity,
the spectral function in the double well problem of dissipative
quantum mechanics and part of the
space dependent succeptibility in the Kondo model .

\Date{03/96}

\newsec{Introduction.}

One dimensional
quantum impurity problems arise in diverse areas of  solid-state
physics. Of
recent interest
are the tunneling through a point contact
in the fractional quantum Hall effect \moon,
the Kondo problem \Kondo\ describing electrons interacting with
dilute
impurities in a metal, and  the dynamics  of double or  multiwell
systems coupled to an ohmic bath in  the context of
dissipative quantum mechanics \Leggett,\ref\ChRu{S. Chakravary, J.
Rudnick,
Phys. Rev. Lett. 75 (1995) 501.} .
A common
feature to all these models is that they can be reduced to a model
described by massless excitations in the bulk interacting
with an impurity at the boundary.
The absence of a mass gap leads to a power law behaviour for the
current correlators in both the ultra-violet and the
infra-red regime .  The cross-over between these two regimes is
non-trivial
because of the renormalisation group flow induced by the impurity.
A standard approach to these systems would be to use perturbation
theory but it fails to capture all the physics, and  new methods
are needed.

Beside (largely numerical) renormalization group
approaches, another  possibility for progress is provided  by  the
integrability of
some of these systems.  Albeit not all of them are integrable,
surprisingly, many are, and exact results can be obtained.
The oldest  such results concern the
Toulouse limit \AYH\ of the Kondo model,
which is a special point in the parameter space of the anisotropic
model that is equivalent to free fermions.
Other  examples include  the thermodynamic properties of the
Kondo problem \thegoodguys\  which were computed using the
Bethe ansatz, and more recently the tunneling
through a point contact in the $\nu=1/3$ quantum Hall effect
 \FLS, where thermodynamic and as well
some static transport properties were computed  by a combination
of
Bethe ansatz and Boltzmann equations.
Apart from the latter solution,  transport properties as well
as space and time-dependent properties
are difficult to obtain because they require knowledge of correlation
functions.

Impurity problems of physical interest have  a very simple
bulk hamiltonian, typically a free boson. All the difficulty lies
in the impurity interaction. In the basis where the
bulk problem is simple  this interaction
is difficult to hanlde:  in the classical limit, plane
waves are scattered into
very complicated wave packets by the impurity.
If the theory is integrable, there is
however another basis onto which the impurity has a simple effect:
 in the classical limit, there are
some special wave packets that scatter simply at the impurity \FSW.
The price to pay  of course is that in this new basis the bulk
hamiltonian looks
more  complicated. Typically, this will lead us to describe a simple
free
boson as a gas of massless quasiparticles (solitons, antisolitons and
breathers) with non trivial scattering! The net result however is
that
both bulk and impurity are now manageable, and physical properties
can be computed.

Let us now recall that integrable massive theories, eg the bulk
sine-Gordon model, can be described as a gas of quasiparticles with
factorized scattering. A convenient formalism is to introduce
creation and annihilation operators for
these quasiparticles, like one would normally do for say a theory  of
free Fermions.
If we denote by $Z^*_{\epsilon}(\theta)$ (
$Z^{\epsilon}(\theta)$ ) the creation (anihilation) operator of a
quasiparticle of type $\epsilon$,
the bulk interaction is encoded in the following Fadeev-Zamolodchikov
relations~:
\eqn\fadzamo{\eqalign{
Z^{\epsilon_1}(\theta_1)Z^{\epsilon_2}(\theta_2)&=
S^{\epsilon_1\epsilon_2}_{
\epsilon_1'\epsilon_2'}(\theta_1-\theta_2) Z^{\epsilon_2'}(\theta_2)
Z^{\epsilon_1'}(\theta_1)\cr
Z^*_{\epsilon_1}(\theta_1)Z^*_{\epsilon_2}(\theta_2)&=
S^{\epsilon_1'\epsilon_2'}_{
\epsilon_1\epsilon_2}(\theta_1-\theta_2)Z^*_{\epsilon_2'}(\theta_2)
Z^*_{\epsilon_1'}(\theta_1)  \cr
Z^{\epsilon_1}(\theta_1)Z^*_{\epsilon_2}(\theta_2)&=
S_{\epsilon_2\epsilon_1'}^{
\epsilon_2'\epsilon_1}(\theta_1-\theta_2)
Z^*_{\epsilon_2'}  Z^{\epsilon_1'}(\theta_1)+
2\pi\delta_{\epsilon_2}^{\epsilon_1} \delta(\theta_1-\theta_2).
}}
In these expressions, the $\theta_i$'s are rapidity variables. In the
discussion so far  the theory is massive,
and there is a meaningful mass scale $m$ related to the coupling
constant of some bulk perturbation. The energy and the momentum of
quasiparticles can be then parametrized in the form
$E_\epsilon=m_\epsilon \cosh(\theta)$ and $P_\epsilon=m_\epsilon
\sinh(\theta)$,
with $m_\epsilon\propto m$.  For the relations \fadzamo\ to make
sense,
constraints have to be satisfied by the $S$ matrix, in particular it
has to be a solution of the Yang-Baxter equation.

It is then convenient to  think of a massless theory as the
limit of a massive theory when the bulk mass scale $m$ goes to zero;
for instance
one can think of a free boson theory as the limit of the bulk
sine-Gordon model
when the amplitude of the bulk cosine perturbation vanishes\foot{Such
a description requires in particular that the space of fields of the
massive field theory considered as a perturbed conformal field theory
and its massless
ultraviolet limit are identical, which is the case in the
``superrenormalizable'' theories we consider here \ref\AZ{A.
Zamolodchikov, in ``Advanced Studies in Pure Mathematics'' 19 (1989)
641.}.}.  Consider thus a massive theory, and take
the following (massless) limit: $\theta=\theta_0\pm\beta$
and $m\rightarrow 0$ such that $M=me^{\theta_0}/2$ remains
finite.  Excitations split into left and right movers,
and dispersion relations  become~:
\eqn\mass{\eqalign{
E_\epsilon&=P_\epsilon=M_\epsilon e^\beta \ \ \ \ {\rm right movers}
\cr
E_\epsilon&=-P_\epsilon=M_\epsilon e^{-\beta} \ \ {\rm left movers} .
}}
Here $M$ is  an arbitrary  energy scale without physical meaning.
The previous relations \fadzamo\ are then decorated by a
supplemental L or R subscript.  The $S$ matrices for the
interaction between  movers of the same chirality are unchanged,
since
in the initial massive theory  they depended only on the rapidity
difference
\foot{The $S$ matrix depends on the relativistic invariant
$(E_1+E_2)^2-(P_1+P_2)^2$ in the massive case, $E_1/E_2$ in the
massless case,
both functions of the rapidity difference using our
parametrization}, but
the right-left scattering becomes a constant phase (we refer the
reader to \FSW\ for a more
detailled discussion).  This trivial left-right scattering
will lead  in the form factors approach to a left-right  
factorization,
and will simplify matters drastically.

This massless  basis
was not explicitely used
in the original exact approaches  to the Kondo problem. In the
corresponding
works, bare hamiltonians
were used instead, and explicitely diagonalized via the Bethe ansatz.
The point
of view we take here is quite different: we work directly in the
renormalized theory which is assumed to be factorizable \ref\ZZ{A. B.
Zamolodchikov,
Al. B. Zamolodchikov, Ann. Phys. 120 (1979) 253.} (this following
of course from the integrability of the bare hamiltonian). This makes
a tremendous difference when dealing with matrix elements of fields.
While
their computation from the bare theory is extremely arduous,
they can be obtained rather easily in the context of the
factorized renormalized theory by using an axiomatic approach.
The massless basis
has proven very convenient
in recent works \ref\PKon{
P. Fendley, Phys. Rev. Lett. 71 (1993) 2485.},
in particular for what concerns static transport properties, like
the DC conductance and the  DC noise \FLS , \FLSnoise\ in the problem
of tunneling between Hall edge states. Here, following the idea 
described in the pioneering work of 
\ref\DMS{G. Delfino, G. Mussardo,
P. Simonetti, Phys. Rev. D51, (1995) 6620.}, we show how this basis
also permits the computation of time (or space) dependent properties.

The natural approach to get  correlations is to use
matrix elements in the basis \fadzamo, the so-called   form-factors.
  In the massive theory,
these matrix elements are computed using a set of ``axioms"
\Kar,\Smir, \Mussi, generalizing Watson's theorem \W.  Form-factors
\foot{These are actually
called ``generalized form-factors'' since no order of the rapidities
is prescribed} of an operator ${\cal O}$  in a bulk theory are
defined as~:
\eqn\ffacteg{
f(\theta_1,...,\theta_n)_{\epsilon_1,...,\epsilon_n}=
<0|{\cal O}(0,0)\ Z^{*}_{\epsilon_1}(\theta_1)\ldots
Z^{*}_{\epsilon_n}(\theta_n)|0>
}
where $|0>$ is the ground state, and their determination  results
form the following axioms~:
\eqn\axiomi{\eqalign{
f(\theta_1,...&,\theta_i,\theta_{i+1},...,\theta_n)_{\epsilon_1,...,
\epsilon_i,\epsilon_{i+1},...,\epsilon_n}
S^{\epsilon_i,\epsilon_{i+1}}_{\epsilon_i',\epsilon_{i+1}'}
(\theta_i-\theta_{i+1})\cr =&
f(\theta_1,...,\theta_{i+1},\theta_i,...,\theta_n)_{\epsilon_1,...,
\epsilon_{i+1},\epsilon_i,...,\epsilon_n} .\cr
}}
which is a consequence of \fadzamo , and~:
\eqn\axiomii{
f(\theta_1,...,\theta_n+2\pi i)_{\epsilon_1,...,\epsilon_n} =
f(\theta_n,\theta_1,...,\theta_n+2\pi
i)_{\epsilon_n,\epsilon_1,...,\epsilon_{n-1}} ,
}
which is a generalisation of the two particle form factor monodromy
equations \Smir .
The $S$ matrix in the first relation has anihilation poles and this
will induce similar poles in the form factors.  Thus
when $\theta_n=\theta_{n-1}+i\pi$ (which correspond to the mass
thresholds in the two particles case for example) we have~:
\eqn\iomiii{\eqalign{
i &\ res \ f(\theta_1,...,\theta_n)_{\epsilon_1,...,\epsilon_n} =
f(\theta_1,...,\theta_{n-2})_{\epsilon_1',...,\epsilon_{n-2}'}
C_{\epsilon_n,\epsilon_{n-1}'}
\times \cr &
\left(
1-S_{\tau_1\epsilon_1}^{\epsilon_{n-1}'\epsilon_1'}
(\theta_{n-1}-\theta_1)\cdots
S_{\tau_{n-3}\epsilon_{n-3}}^{\tau_{n-4}\epsilon_{n-3}'}
(\theta_{n-1}-\theta_{n-3})
S_{\epsilon_{n-1}\epsilon_{n-2}}^{\tau_{n-3}\epsilon_{n-2}'}
(\theta_{n-1}-\theta_{n-2})
\right).
}}
Also, if the theory contains
bound states, there are further relations involving the corresponding
residues
in the $S$ matrix.  For example, if there is a bound-state at
$\theta_n=\theta_{n-1}+i u$, corresponding to a diagram depicted
in figure 1.
\fig{Bound state diagram.}{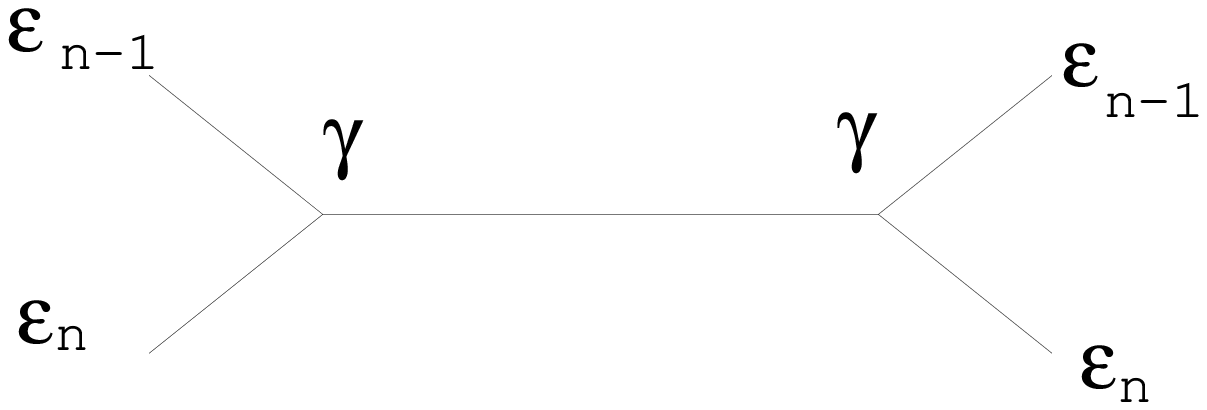}{5cm}
\figlabel\tabb
Then, the corresponding relation for the form factor is~:
\eqn\respo{
i \ res \ f(\theta_1,...,\theta_n)_{\epsilon_1,...,\epsilon_n} =
a_\gamma (-1)^{(2\epsilon_n+1)/2} C_{\epsilon_{n-1},\epsilon_n}
f(\theta_1,...,\theta_{n-1}-iu)_{\epsilon_1,...,\epsilon_{n-2},\gamma},
}
and here $a_\gamma$ correspond to the square root of the
corresponding
residue in the $S$ matrix for this process.

Solving  this set of equations is the most convenient method  to
obtain
the matrix elements (although other methods based on
lattice regularizations might also be applicable \Korbook).  This  is
still a  difficult task, but results are available for the
sine-Gordon \Smir\ and
the sinh-Gordon \Mussi\ models, which is all what
we need in this paper.  Form-factors for the massless theory will
be defined simply by taking the foregoing limit of massive
form-factors (trying to formulate  axioms  for a massless theory per
se
might lead to some ambiguities).
Our strategy will then be to compute correlators simply by
decomposing
on intermediate states and using the exact matrix elements. This
might sound
a priori hopeless since there are actually an infinite sum of
relevant terms
for operators of physical interest; however, we shall see that in
many cases,
the problem simplifies drastically.

The paper is organized   as follows. In section 2 we  introduce the
models
and the quantities
of physical interest which we want to compute.  In section 3, the
technique
of form-factors is introduced with the example of the sinh-Gordon
model. Although it is not of immediate physical relevance, this model
is very simple
since it has  a single quasi-particle excitation, and appears useful
pedagogically.
In section 4, we introduce similarly the form-factors for the
sine-Gordon model
and compute the
frequency dependent conductance in the $\nu=1/3$ quantum Hall
effect. In section 5, further applications are discussed; in
particular we compute
the spectral function in dissipative quantum mechanics and the
uniform part of the space
dependent succeptibility in the anisotropic Kondo model
related to the screening-cloud problem.

\newsec{The models.}

The impurity problems we shall discuss here can all be mapped
onto a hamiltonian of the form\foot{In all what follows
we set $e=h=1$.}~:
\eqn\hamgen{
H={1\over 2} \int_{-\infty}^0 dx \
[8\pi g \Pi^2+{1\over 8\pi g}(\partial_x\phi)^2] + {\cal B}.
}
where ${\cal B}$ is a problem dependent boundary interaction and the
fields are defined
on the negative half line.  The geometry of the
problem is shown in figure 2.
\fig{Geometry of the problem.}{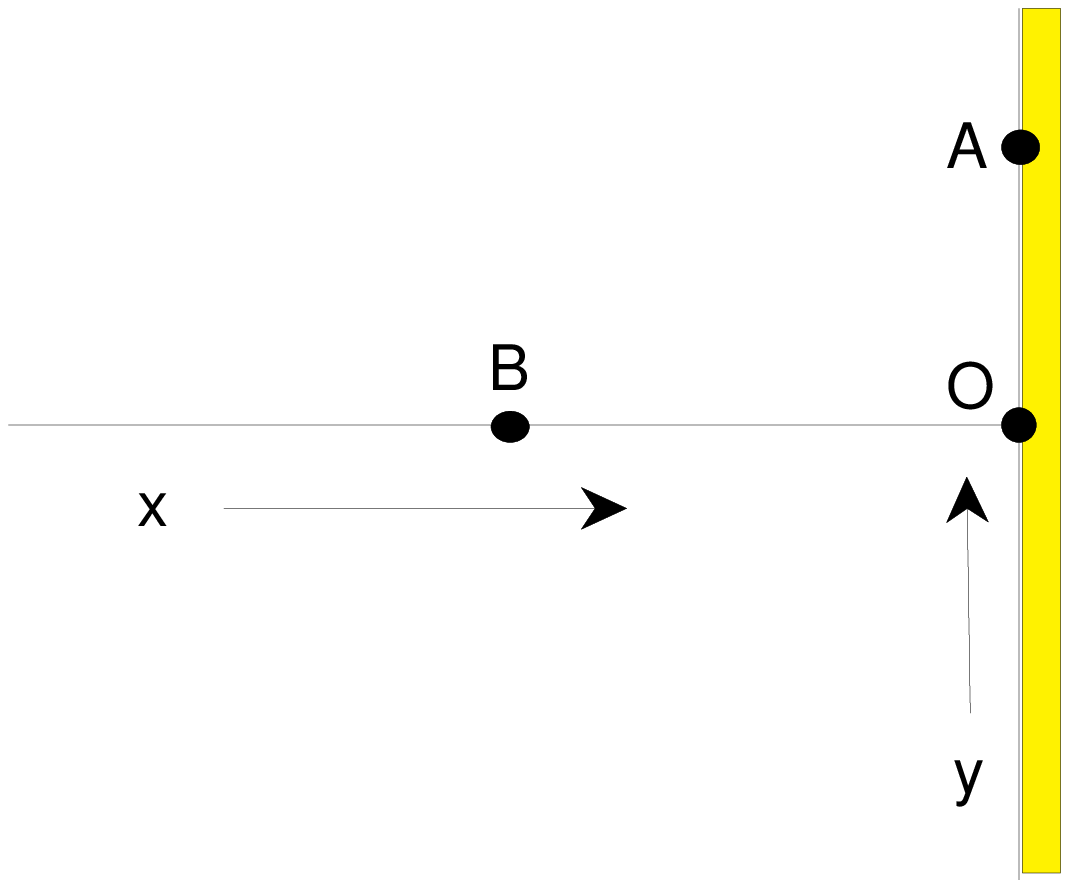}{6cm}
\figlabel\tabb
The method we present is quite
general and in this section we proceed to an enumeration of the
models we study and the corresponding quantities we want to
compute~:

\subsec{Sinh-Gordon model.}

In this model, the boundary interaction is of the form~:
\eqn\sih{
{\cal B}=\lambda \ \cosh {1\over 2}\phi(x=0,t).
}
Although we have no physical application for this model, it is
much simpler than the models we will explore subsequently.
The spectrum of massless excitations consist of only one particle
and the form factors are much simpler than those used later
in the sine-Gordon model.
The quantities we will compute are the current-current correlation
and the equivalent of the conductance in the Hall model.

Our motivation for its study is to develop, in a simple fashion,
the techniques used later in the real physical problems.  We
will show in a more lengthy manner how to obtain the current
correlation functions and how the boundary affects these
correlations.  All features found in the other models are found in
this one as well so it is a very good exercise.

\subsec{$\nu=1/3$ Hall effect.}

To start,
let us recall  the relation between the  physical
sine-Gordon model with an impurity  and
the problem in the half space. Let us start with the hamiltonian~:
\eqn\remi{H={1\over 2}\int_{-\infty}^\infty dx \
[8\pi \nu \Pi^2+{1\over 8\pi \nu} (\partial_x\varphi)^2
]+\lambda \delta(x) \cos(\varphi_L-\varphi_R),}
where the L and R components depend on $x,t$ as
$\varphi_L(x+t),\varphi_R(x-t)$.  The bulk part of
this hamiltonian physically
describes the low energy edge degrees
of freedom in the quantum Hall effect at filling fraction
$\nu={1\over 2s+1}$.
When this Hall sample is subjected to a point contact constriction,
transfer of fractionally charged excitations is possible.  For
generic
filling fraction, many types of quasi-particles contributes as
relevant
charge transfer.  The $\nu=1/3$ is peculiar in that only the $Q=e/3$
charged
laughlin quasi-particle is relevant \moon .  This
is described by the impurity term in \remi .
As discussed in \FLS\ in order to map this to a boundary
problem, it is convenient to proceed
in two steps. First introduce~:
\eqn\remii{\eqalign{\phi^e(x+t)={1\over\sqrt{2}}\left[\varphi_L(x,t)+
\varphi_R(-x,t)\right]\cr
\phi^o(x+t)={1\over\sqrt{2}}
\left[\varphi_L(x,t)-\varphi_R(-x,t)\right] \cr}}
which are both left moving. It is clear that the interaction term
does not affect the even field, which therefore remains free.
As for the odd term, it can be mapped onto a boundary problem
as follows. Define~:
\eqn\remiii{\eqalign{\phi^o_L(x,t)&=\sqrt{2}\phi^o(x+t), \ \ x<0, \cr
\phi^o_R(x,t)&=\sqrt{2}\phi^o(-x+t), \ x<0.\cr}}
The odd hamiltonian then reads~:
\eqn\remiv{H={1\over 2}\int_{-\infty}^0
[8\pi g (\Pi^o)^2+{1\over 8\pi g}(\partial_x\phi^o)^2]
+\lambda \delta(x) \cos{1\over 2}\phi^o,}
and in the following we will write  $\phi\equiv \phi^o$ and $g$
instead of
$\nu$.  Thus, for
this problem, ${\cal B}=\lambda  \cos{1\over 2}\phi(x=0,t)$.

The quantity of interest in this case is
the AC conductance at vanishing temperature.
A standard way of representing it is through the Kubo formula~:
\eqn\kubozero{G(\omega_M)=-{1\over 8\pi \omega_M L^2}\int_{-L}^L
dx \ dx'\int_{-\infty}^\infty dy \ e^{i\omega_M y}<j(x,y)
j(x',0)>,}
where $\omega_M$ is a Matsubara frequency, $y$ is imaginary time,
$y=it$.
One gets back to real physical frequencies by letting
$\omega_M=-i\omega$.
In \kubozero , $j$ is the physical current in the unfolded system,
$j=\partial_t(\varphi_L-\varphi_R)$.  Without impurity, the AC
conductance of the Luttinger liquid is frequency independent, $G=g$.
When adding the impurity, it becomes $G={g\over 2}+\Delta G$. After
some
simple manipulations using the folding, one finds~:
\eqn\kubozeri{\eqalign{\Delta G(\omega_M)&={1\over 8\pi \omega_M
L^2}\int_{-L}^0
dxdx'\int
_{-\infty}^\infty dy e^{i\omega_M
y} \cr &\left[<\partial_z\phi(x,y)\partial_{\bar{z}'
}\phi(x',0)>
+<\partial_{\bar{z}}\phi(x,y)\partial_{z'}\phi(x',0)>\right],\cr}}
where $z=x+iy$.

\subsec{Dissipative quantum mechanics and Kondo model.}

The second model we will be interested in is the anisotropic Kondo
model~:
\eqn\hamilgen{
H={1\over 2} \int_{-\infty}^0 dx  [8\pi g \Pi^2+{1\over 8\pi
g}(\partial_x\phi)^2]
+ \lambda \ ( S_+ e^{i\phi(0)/2}+S_- e^{-i\phi(0)/2}).
}
It can be related to dissipative quantum mechanics \Leggett\ where it
describes the dynamics of a double well system subjected to
Ohmic dissipation.  The hamiltonian for such a system is given by~:
\eqn\spinbos{
H=-{\hbar \Delta\over 2}\sigma_x+{\hbar \epsilon\over 2}\sigma_z+
\sum_\alpha \left[
{p_\alpha^2\over 2 m_\alpha}+{1\over 2} m_\alpha
\omega_\alpha^2\left(
x_\alpha-{C_\alpha\over m_\alpha \omega_\alpha^2}\sigma_z\right)^2
\right].
}
In this expression, the Pauli matrices $\sigma_x,\sigma_z$ act on the
two dimensional space of states.  $\Delta$ is a tunnelling matrix
element
and $\epsilon$ denotes a bias between the two states.  Dissipation
comes
from the coupling of this two states system to an environment of
oscillators described by the last part of the hamiltonian.  The
effect of these oscillators (with mass, frequencies and
coupling $m_\alpha$, $\omega_\alpha$, $C_\alpha$) can be entirely
encoded in the environment spectral function~:
\eqn\specfunc{
J(\omega)={\pi\over 2}\sum_\alpha \left( {C_\alpha^2\over
m_\alpha \omega_\alpha}\right) \delta(\omega-\omega_\alpha).
}
It was shown
\Leggett\ that
this system can be mapped, in the so called ohmic dissipation case,
$J(\omega)=2\pi g \omega$, to
an anisotropic Kondo model.  The conduction electrons in the latter
play
the role of dissipation and the $z$ value of the spin is associated
with the
two states.  In this paper we will work at zero bias (the bias would
be equivalent
to a magnetic field in \hamilgen).  The hopping
$\Delta$ is related the the strength of the impurity $\lambda$ the
precise relation
being given in \Leggett .  This system has numerous physical
applications \ref\weiss{
U. Weiss, {\it Quantum Dissipative Systems}, World Scientific,
Singapore, 1993.}.

The quantity of physical interest in the two state system is the
effect of the bath on the tunneling between the two (degenerate
minima). It is conveniently encoded into the correlator ~:
\eqn\respomeg{
C(t)={1\over 2}<[S_z(t),S_z(0)]>.
}
It describes the probability to be in a state $S_z(t)$ given that
the system was in state $S_z(0)$ at $t=0$.
We will show how this is related to a current correlation and
can therefore be addressed by the form factor approach.

Another problem of interest is the screening cloud problem in the
Kondo model.  The three dimensional Kondo model can be reduced
to one dimension by restricting to the s wave~:
\eqn\swar{
\psi({\bf r})={1\over 2\sqrt{2}\pi r} [e^{-ik_F r} \psi_L(r)-e^{ik_F
r}\psi_R(r)]+ \cdots
}
with the $\cdots$ denoting higher harmonics.   In this language the
Hamiltonian
is given by~:
\eqn\konhamil{
H=v_F \int_0^\infty dr \ \psi^\dagger_L(r) {id\over dr}\psi_L(r)-
\psi^\dagger_R(r) {id\over dr}\psi_R(r)
}
and the boundary interaction is~:
\eqn\binter{
{\cal B}=\lambda \psi_L^\dagger(0){{\bf\sigma}\over 2}\psi_L(0)\cdot
{\bf S}_{imp}.
}

The screening cloud problem can be addressed by computing the local
succeptibility which in 3 d is given by~:
\eqn\losucc{
\chi({\bf r})=<\psi^\dagger ({\bf r}){\sigma_z\over 2}\psi({\bf r})
\int dt \ S_{tot}^z>,
}
with~:
\eqn\stot{
S_{tot}^z=S_{imp}^z+\int d^3 r \ \psi^\dagger({\bf r}){\sigma_z\over
2}\psi({\bf r}).
}
Here $S_{imp}$ correspond to the impurity spin and the second part is
the electron spin (the spin indices of the fermions are contracted
with
the Pauli matrix).  When reducing these quantities to one dimension,
we
find following \Affleck\ a uniform and $2k_F$ part written~:
\eqn\undsucc{
\chi(r)={1\over 8 \pi^2 r^2}[\chi_{un}+2 \chi_{2k_F} \cos(2 k_F r)]
}
with~:
\eqn\det{\eqalign{
\chi_{un}&=<[\psi^\dagger_L(r){\sigma_z\over 2}\psi_L(r)+
\psi^\dagger_R(r){\sigma_z\over 2}\psi_R(r)] \int dt S^z_{tot}> \cr &
\chi_{2k_F}=<[\psi^\dagger_L(r){\sigma_z\over 2}\psi_R(r)+
\psi^\dagger_R(r){\sigma_z\over 2}\psi_L(r)]\int dt S^z_{tot}>
}}
and the total spin now given by~:
\eqn\sstot{
{\bf S}_{tot}=S_{imp}^z+{1\over 2\pi}\int_0^\infty  dr \
[\psi^\dagger_L(r){\sigma\over
2}\psi_L(r)+\psi^\dagger_R(r){\sigma\over 2}
\psi_R(r)].
}

After  having  established the quantities  we want to compute,
we can  bosonise\Schotte. Two bosonic
fields are necessary: one associated with charge and one with spin.
The charge field
decouples completely and only the spin charge has interaction at the
boundary.  The action for the spin field is the of the form \hamilgen
{}.  The $S_z$ term in the action has been eliminated by a unitary
rotation of the hamiltonian (which is unity at $g=1$, the
isotropic model).
In terms of this bosonic field, the uniform part of the
succeptibility at $g=1$
takes the simple form~:
\eqn\bosucc{
\chi_{un}(r)={1\over 2} <\partial_r \phi(r) \int dt \ S_{tot}^z>
}
with now\foot{This expression has to be considered with some caution
but will
be enough for the screening cloud computations of section 5.}~:
\eqn\bostot{
S_{tot}^z=S_{imp}^z+ {1\over 4\pi}\int_{-\infty}^0 dx
\partial_x\phi(x).
}
We will show how to compute this uniform part at zero temperature.

The problems are posed and
our task is then to compute these correlation functions. This is
a long story, and to explain the method we start by considering
as a toy model the boundary sinh-gordon theory.

\newsec{Formalism: The sinh-Gordon model.}

\subsec{The bulk current-current correlators.}

In most of this paper, we shall work  in Euclidian space with $x,y$
coordinates. Imaginary
time is at first considered as running along $x$.
We consider the sinh-Gordon model with action~:
\eqn\shgoract{S={1\over 16\pi g}\int_{-\infty}^\infty
dxdy\left[\left(\partial_x
\phi\right)^2
+ \left(\partial_y\phi\right)^2+\Lambda\cosh \phi\right].}
This is a massive theory which is integrable; the conformal weights
of the perturbing operator are $h=\bar{h}=-g$.  The spectrum is very
simple and
consists of a  single particle of masse $m$,  and S matrix~:
\eqn\smat{S(\theta)={\tanh{1\over 2}\left(\theta-i {\pi B\over
2}\right)\over
\tanh{1\over 2}\left(\theta+i {\pi B\over 2}\right)},}
where~:
$$
B={2g\over 1+g}
$$
and $\theta$ is the usual rapidity, $P=m\sinh\theta$,
$E=m\cosh\theta$.
Recall the duality of the S matrix in $B\to 2-B$, i.e. in $g\to 1/g$.

Consider the massless limit where $\Lambda\to 0$. In that limit,
the current correlators are trivial and are given by~:
\eqn\corr{\eqalign{<\partial_z\phi(z,\bar{z})
\partial_{z'}\phi(z',\bar{z}')>=
-{2g\over (z-z')^2}\cr
<\partial_{\bar{z}}\phi(z,\bar{z})\partial_{\bar{z}'}
\phi(z',\bar{z}') >=
-{2g\over (\bar{z}-\bar{z}')^2}.\cr}}
On the other hand, we can formally describe this limit still using a
scattering theory. As described in the introduction, this is done by
writting
$\theta=\pm(\theta_0+\beta)$ and taking simultaneously
$\theta_0\to\infty$ and $m\to 0$ with $me^{\theta_0}/2\to M$, $M$
finite. In that case, the spectrum separates into Right  and Left
movers
with respectively $E=P=M e^\beta$ and $E=-P=M e^\beta$.
The scattering of R and L movers is still given by \smat\ where
$\theta\to\beta$. The RL and LR scattering becomes a simple phase,
$e^{\mp i\pi B/2}$. This phase will turn out to cancel out at the end
of
all computations, but is confusing to
keep along. We just set it equal to unity in the following, that is
we consider all L and R quantities as commuting.

In this new description of the massless theory, we will need form
factors in order to compute \corr .
The form factors of the massive sinh-Gordon theory are well known
\ref\Giu{A.Fring, G. Mussardo, P. Simonetti, Nucl. Phys. {\bf B}393
(1993) 413; G. Mussardo, A. Koubek, Phys. Lett. B.311 (1993), 193;
G. Mussardo, A. Koubek, P. Simonetti, Int. J. Mod. Phys.
 A 9 (1994), 3307. }. We will only use
the form factors of the fundamental field $\phi$ in what follows.
By taking the massless limit of the formulas in \Giu, it is easy to
check that $\phi$ can alter only the right or left content of states;
in other words, matrix elements of $\phi$
between states which have different content both in the
left and right sectors vanish.

Our conventions are conveniently summarized by giving
the one particle form factor of the sinh-Gordon field~:
\eqn\conv{\eqalign{<0|\phi(x,y)|\beta>_R&= \mu \exp\left[ Me^\beta
(x+iy)\right]\cr
<0|\phi(x,y)|\beta>_L&=\mu \exp\left[Me^\beta (x-iy)\right].\cr}}
For  the field $\phi$, form factors with an even number of particles
vanish.
This is because $\phi$ as well as the creation operators
of the sinh-gordon particle are  odd under the $Z_2$ symmetry
$\phi\to-\phi$.
In the following we will use the notation~:
\eqn\nnot{f(\beta_1,\ldots,\beta_{2n+1})=
<0|\phi|\beta_1,\ldots,\beta_{2n+1}>_{R,\ldots,R},}
with the normalization of asymptotic states
$_R<\beta|\beta'>_R=2\pi\delta(\beta
-\beta')$. Here, $f$ depends on $g$,
but we do not indicate it explicitely for simplicity.
We have the following properties~:
\eqn\propp{\eqalign{<0|\phi|\beta_1,\ldots,\beta_{2n+1}>_{R,\ldots,R}= 
\left(<0|\phi|\beta_1,\ldots,\beta_{2n+1}>_{L,\ldots,L}\right)^* \cr
<0|\phi|\beta_1,\ldots,\beta_{2n+1}>_{R,\ldots,R}=
<0|\phi|\beta_{2n+1},\ldots,\beta_1>_{L,\ldots,L},\cr}}
These form factors are expressed as follows \Giu . Introduce~:
\smallskip
\eqn\fmin{f_{min}(\beta)={\cal N}\exp\left\{8\int_0^\infty {dx\over
x}
{\sinh ( {xB\over 4})\sinh [{x(2-B)\over 4}]\sinh ({x\over 2})\over
 \sinh^2(x)}\sin^2 \left[
{x(i\pi-\beta)\over 2\pi}\right]
\right\},}
where~:
$$
{\cal N}= \exp\left[-4\int_0^\infty {dx\over x}
{\sinh ({xB\over 4})\sinh [{x(2-B)\over 4}]\sinh ({x\over 2})\over
 \sinh^2(x)}\right]
$$
Then,
\eqn\ffexpr{f(\beta_1,\ldots,\beta_{2n+1})=\mu
\left({4\sin {\pi B\over 2} \over F_{min}(i\pi,B)}\right)^n
\sigma_{2n+1}^{(2n+1)}P_{2n+1}(x_1,\ldots,x_{2n+1})
\prod_{i<j}{f_{min}(\beta_i-\beta_j) \over x_i+x_j},}
where we introduced $x\equiv e^\beta$ and the $\sigma$'s are the
basic
symmetric polynomials~:
$$
\sigma_p=\sum_{i_1<i_2<\cdots <i_p} x_{i_1} x_{i_2}\cdots x_{i_p},
$$
with the convention $\sigma_0=1$ and $\sigma_p=0$ if $p$ is greater
than the number of variables.
The $P_{2n+1}$'s are symmetric polynomials, which can be obtained
by solving LSZ recursion relations. The first ones read~:
\eqn\poly{\eqalign{P_3(x_1,\ldots,x_3)=&1\cr
P_5(x_1,\ldots,x_5)=&\sigma_2\sigma_3-c_1^2\sigma_5\cr
P_7(x_1,\ldots,x_7)=&\sigma_2\sigma_3\sigma_4\sigma_5-
c_1^2(\sigma_4\sigma_5^2
+\sigma_1\sigma_2\sigma_5\sigma_6+
\sigma_2^2\sigma_3-c_1^2\sigma_2\sigma_5
)\cr
&-c_2(\sigma_1\sigma_6\sigma_7+\sigma_1\sigma_2\sigma_4\sigma_7)+
\sigma_3\sigma_5
\sigma_6)+c_1c_2^2\sigma_7^2.\cr}}
with $c_1=2\cos \pi B/2$, $c_2=1-c_1^2$. Observe that except for the
overall
normalization $\mu(g)$, these expressions are invariant in the
duality transformation $g\to {1\over g}$.

In \conv, $\mu$ is an overall normalization for the form-factors.
It is usually chosen by reference to the IR limit. However, we will
require
that the result \corr\ be recovered, and this  involves a more
complex
computation. Using form factors, this two point function expands,
assuming $Re\ z<Re\ z'$, as~:
\eqn\ffnorm{\eqalign{<0|\partial_z\phi(z,\bar{z})\partial_{z'}
\phi(z',\bar{z}')|0>=
-\sum_{n=0}^\infty \int {d\beta_1\ldots d\beta_{2n+1}\over
(2\pi)^{2n+1}(2n+1)!}
M^2 \left(e^{\beta_1}+\ldots+e^{\beta_{2n+1}}\right)^2\cr
\exp
\left[M(z-z')\left(e^{\beta_1}+\ldots+e^{\beta_{2n+1}})\right)\right]
|f(\beta_1,\ldots,\beta_{2n+1})|^2.\cr}}
Now, by relativistic invariance, all the form factors depend
only on differences of rapidities. Setting $M(z-z')\equiv
e^{\beta_0}$,
(where $\beta_0$ will in general be complex), one
can shift all the  $\beta$'s  by $\beta_0$ to factor out, for
any $2n+1$ particle contributions, a factor ${1\over (z-z')^2}$.
Hence,
the form factor expansion gives the result \corr\ provided $N$ is
chosen
such that
\eqn\niden{\sum_{n=0}^\infty I_{2n+1}=2g,}
where
\eqn\deff{
I_{2n+1}= \int {d\beta_1\ldots d\beta_{2n+1}\over
(2\pi)^{2n+1}(2n+1)!}
 \left(e^{\beta_1}+\ldots+e^{\beta_{2n+1}}\right)^2
e^{-(e^{\beta_1}+\ldots+e^{\beta_{2n+1}})}
|f(\beta_1,\ldots,\beta_{2n+1})|^2.}
In practice, this sum cannot be computed analytically, but it
can be easily evaluated numerically. The
convergence is extremely fast with $n$, and for most practical
purposes, the
consideration of up to five particles is enough to get
correct results up to $10^{-4}$.  Similar convergence properties
were observed in \ref\cardy{J. Cardy, G. Mussardo, Nucl. Phys. 
B 410 (1993), 451 }.

It must be emphasized that this result
is very peculiar to the current operator. For most other chiral
operators, the correct $(z-z')$ dependence involves a non trivial
anomalous dimension, instead of the naive engineering dimension.
Hence,
this dependence  is not obtained term by term, as observed here,
but rather once the whole series is summed up. Truncating the series
to
any finite $n$ does not, in such cases, give reliable results all the
way
from short to large distances. The current is therefore  an extremely
favorable
case, as would be the stress tensor, and we are
fortunate it  has a lot of physical meaning.

\subsec{Current current correlators with a boundary}

Having fixed the form-factors normalization, let us now consider the
theory with a boundary.  The geometry of the
problem was illustrated in figure 2, where the boundary stands at
$x=0$ and runs
parallel to the $y=it$ axis. The action is now~:
\eqn\bdract{S={1\over 16\pi g}\int_{-\infty}^0
dx\int_{-\infty}^\infty dy
\left[
\left(\partial_x\phi\right)^2
+ \left(\partial_y\phi\right)^2+\Lambda\cosh \phi\right]+\lambda
\int_{-\infty}^\infty dy \cosh{1\over 2}\phi(x=0,y).}
This model is also integrable for any choice of $\Lambda,\lambda$.
The boundary dimension of the
perturbing operator is $x=-g$.
We can in particular take the limit where $\Lambda\to 0$ while
$\lambda$
remains finite.
It then describes a theory which is conformal invariant in the bulk
but has a boundary interaction that breaks this invariance and
induces a flow
from Neumann boundary conditions at small $\lambda$ to Dirichlet
boundary
conditions at large $\lambda$. As discussed in \FSW ,  the boundary
interaction is
characterized by an energy scale,
which one can represent as $T_B=Me^{\beta_B}$. $T_B$ is related
with the bare coupling in the action \bdract\ by $\lambda\propto
T_B^{1+g}$.
In the following, since obviously changes of $M$ (which is not a
physical
scale)
can be absorbed in rapidity shifts, we set $M=1$. The effect of the
boundary is
then expressed by the  reflection matrix~:
\eqn\rmat{R(\beta)=\tanh \left[{\beta\over 2}-i{\pi\over 4}\right].}

In the picture where imaginary time
is along $x$, the effect of the boundary is represented
by a boundary state. Following \GZ\ we can represent
it in terms of the boundary scattering matrix~:
\eqn\Bform{|B>=\exp\left[\int_{-\infty}^\infty {d\beta\over 2\pi}
K(\beta_B
-\beta)Z^*_L(\beta)
Z^*_R(\beta)\right]|0>.}
In this formula, $Z^*$ denote the  Zamolodchikov Fateev creation
operators, $K$ is related with the reflection matrix by~:
\eqn\krrel{K(\beta)=R\left({i\pi\over 2}-\beta\right)=-\tanh
{\beta\over 2}.}
One can expand the boundary state into the convenient form~:
\eqn\Bformconv{|B>=\sum_{n=0}^\infty \int_{0<\beta_1<\ldots<\beta_n}
K(\beta_B-\beta_1)\ldots K(\beta_B-\beta_n)Z^*_L(\beta_1)\ldots
Z^*_
L(\beta_n)
Z^*_R(\beta_1)\ldots Z^*(\beta_n).}
Observe now that by analyticity, the matrix elements of
$\partial_z\phi$
between the ground state and any state with at least one L moving
particle are identically zero.  More generally,  the only
non vanishing matrix elements of $\partial_z\phi$ are those where bra
and ket
have the same L moving part. The same results apply by exchanging
$\partial_z$
with $\partial_{\bar{z}}$ and L with R moving particles. As a result
one gets immediately two of the four current correlators~:
\eqn\trivial{\eqalign{<0|\partial_z\phi(z,\bar{z})
\partial_{z'}\phi(z',\bar{z}')|0>&=
-{2g\over (z-z')^2}\cr
<0|\partial_{\bar{z}}\phi(z,\bar{z})\partial_{\bar{z}'}
\phi(z',\bar{z}')|0>&=-{2g\over (\bar{z}-\bar{z}')^2},\cr}}
which are identical with the ones without a boundary.

The two other correlators are more difficult to
get. Let us consider for instance~:
\eqn\tobf{
<0|\partial_{\bar{z}}\phi(z,\bar{z})
\partial_{z'}\phi(z',\bar{z}')|B>.}
The first non trivial contribution comes
from the two particle term in the expansion of the boundary state~:
\eqn\first{\eqalign{
\int_{-\infty}^\infty {d\beta\over 2\pi}K(\beta_B-\beta) <0|
\partial_{\bar{z}}\phi(z,\bar{z})
\partial_{z'}\phi(z',\bar{z}')Z^*_L(\beta)Z^*_R(\beta)|0>=\cr
\mu^2\int_{-\infty}^\infty {d\beta\over 2\pi} K(\beta_B-\beta)
e^{2\beta}
\exp\left[ e^\beta (\bar{z}+z')\right].\cr}}
More generally, because $|B>$ is a superposition of states with
equal numbers of left and right moving particles, and
$\partial_z\phi$,
respectively $\partial_{\bar{z}}\phi$ act only on R, respectively L,
particles,
the expansion of \tobf\ takes a very simple form~:
\eqn\step{\eqalign{
\sum_{n=0}^\infty \int {d\beta_1\ldots d\beta_{2n+1}\over
(2\pi)^{2n+1}(2n+1)!}
K(\beta_B-\beta_1)\ldots K(\beta_B-\beta_{2n+1})
\left(e^{\beta_1}+\ldots+
e^{\beta_{2n+1}}\right)^2
\cr
\exp
\left[(\bar{z}+z')\left(e^{\beta_1}+\ldots+e^{\beta_{2n+1}}\right)
\right]
|f(\beta_1,\ldots,\beta_{2n+1})|^2.\cr}}
This correlation function depends on the product $e^{\beta_B}
(\bar{z}+z')$. It
is scale invariant in the UV and IR fixed point. These correspond
respectively
to sending $\beta_B$ to $\mp \infty$, that is  the coupling
$\lambda$ in the action
to $0$ or $\infty$, in other words Neumann  or Dirichlet boundary
conditions.
In the first case, $K=1$, in the second, $K=-1$. Comparing with
\ffnorm\ and
\niden\ we find, as expected, that~:
\eqn\res{<0|\partial_{\bar{z}}\phi(z,\bar{z})
\partial_{z'}\phi(z',\bar{z}')|B>=\pm {2g\over(\bar{z}+z')^2},}
for Neumann, respectively Dirichlet boundary conditions.
Although trivial, this result shows that the form factor
expansion is well behaved, and allows us to study the correlator
all the way from the UV to the IR fixed point when there is a
boundary perturbation.  In figures 3 and 4 we show the
one particle (which is independent of $B$)
and three particles contributions.  We observe
that indeed the convergence, by looking at the respective
contributions, is very rapid.
\fig{One particle contribution.}{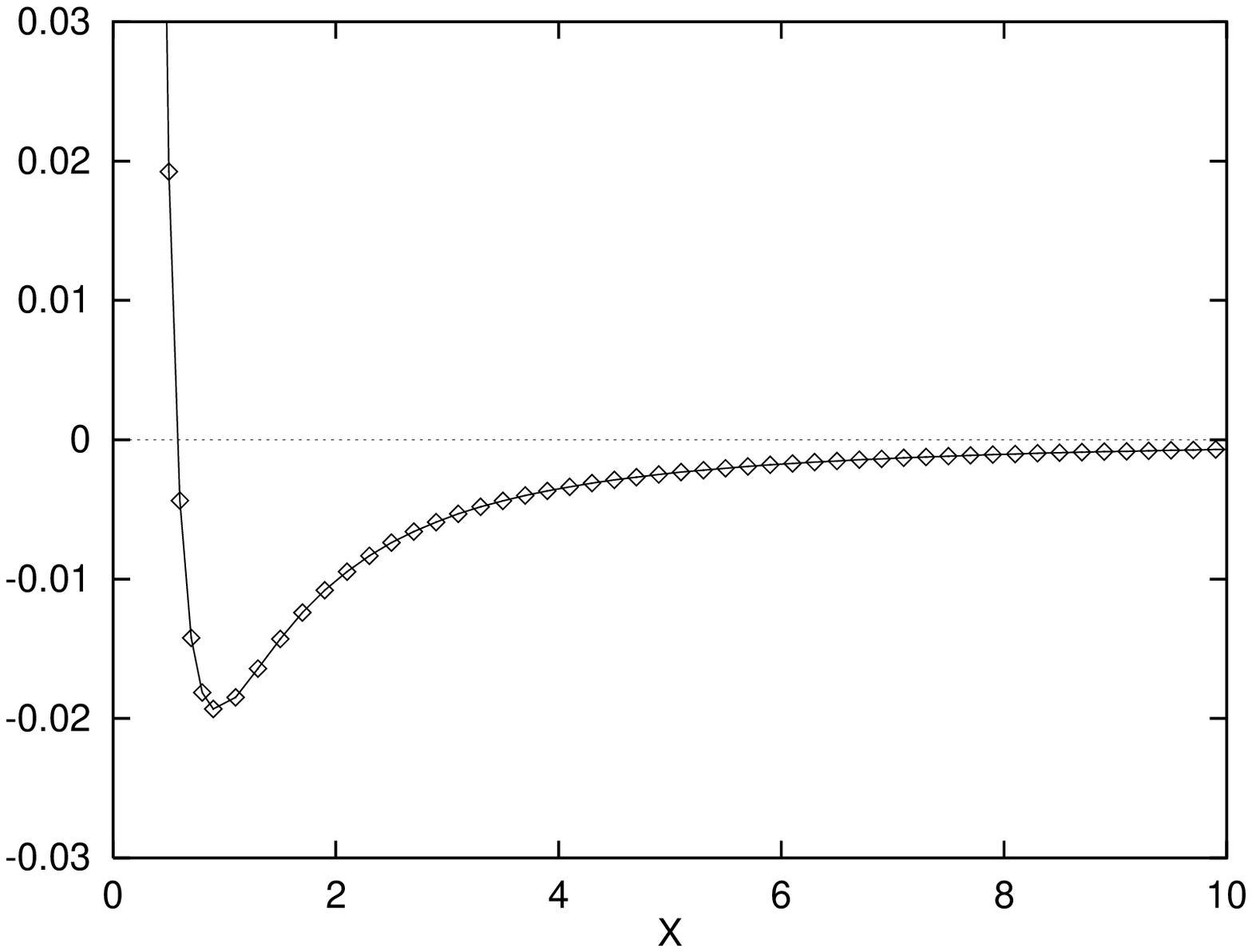}{7cm}
\figlabel\tabb
\fig{Three  particles contribution for $B=1,0.1$.}{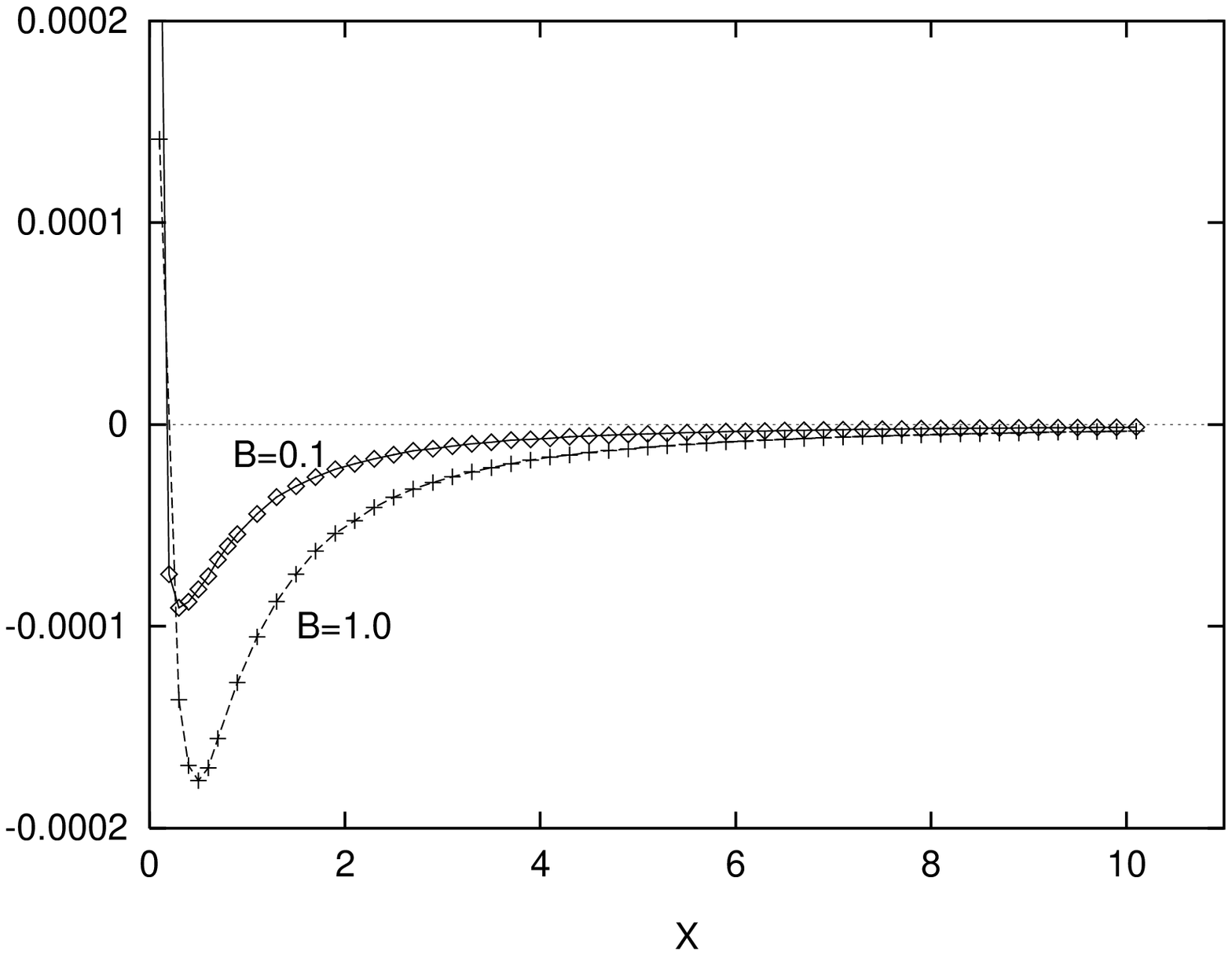}{7cm}
\figlabel\tabb

The only drawback of this expansion is that it  is not suited
for studying the correlation of two operators right at the boundary.
Indeed
in that case, $Re (\bar{z}+z')=0$, and the integrals in \step\ do not
converge.
To solve this problem, we can introduce a modular transformed
picture. We now
consider
the imaginary time as running along the $y$ axis. Now the boundary
is not represented as a state; rather, the whole space of states is
different,
since now we have only a half space to deal with. The asymptotic
states are
not pure L or R moving, but are mixtures. For instance,  one particle
states are~:
\eqn\asym{||\beta>=|\beta>_R+R(\beta)|\beta>_L.}
More generally, asymptotic states are obtained by adding to
$|\beta_1,\ldots,\beta_n>_{R,\ldots,R}$ all combinations with
different
choices of $R$ particles transformed into $L$ particles, via action
of the
boundary. Only the following two terms contribute~:
\eqn\asymr{||\beta_1,\ldots,\beta_n>=
|\beta_1,\ldots,\beta_n>_{R\ldots ,R}
+\ldots+R(\beta_1)\ldots
R(\beta_n)|\beta_n,\ldots,\beta_1>_{L,\ldots,L}+\ldots.}
Although we used the same notation as previously,  different things
are meant by L,R. To make it clear, we now use the conventions~:
\eqn\newconv{\eqalign{<0|\phi(x,y)|\beta>_R
=\mu\exp[e^\beta(-y+ix)]\cr
<0|\phi(x,y)|\beta>_L=\mu\exp[e^\beta(-y-ix)].\cr}}
To keep the notations as uniform as possible, we introduce the new
coordinates~:
\eqn\defdef{w(z)\equiv iz=-y+ix,}
so here R movers depend on $w$, L movers on $\bar{w}$. The
normalization $N$
is of course the same as before, and as before the LL and RR
correlators
do not depend on the boundary interaction. One finds~:
\eqn\trivialbis{\eqalign{<0|\partial_w\phi(w,\bar{w})
\partial_{w'}\phi(w',\bar{w}')|0>&=
-{2g\over (w-w')^2}\cr
<0|\partial_{\bar{w}}\phi(w,\bar{w})\partial_{\bar{w}'}
\phi(w',\bar{w}')|0>&=-{2g\over (\bar{w}-\bar{w}')^2},\cr}}
where we used the fact that $|R(\beta)|^2=1$. When compared with
\trivial,
these
 correlators have an overall minus sign
due to the dimension $h=1,\bar{h}=0$ (resp. $h=0,\bar{h}=1$)
of the operators.

Let us now consider~:
\eqn\tbdbis{<0|\partial_{\bar{w}}\phi(w,\bar{w})\partial_{w'}
\phi(w',\bar{w}')|0>.}
To compute it, we insert a complete set of states which are of the
form \asym. In the
massless case however, since $\partial_w\phi$ is a R operator,
 $\partial_{\bar{w}'}\phi$ a L  operator, the only terms that
contribute are in fact the ones with either all L or all R moving
particles, as written in \asymr. Thus, \tbdbis\ expands simply as~:
\eqn\newexp{\eqalign{-\sum_{n=0}^\infty \int {d\beta_1\ldots
d\beta_{2n+1}\over
(2\pi)^{2n+1}}R(\beta_1-\beta_B)\ldots
R(\beta_{2n+1}-\beta_B)\left(e^{\beta_1}+
\ldots+e^{\beta_{2n+1}}\right)^2\cr
\exp\left[(\bar{w}-w')(e^{\beta_1}+\ldots+e^{\beta_{2n+1}})\right]
|f(\beta_1,\ldots,\beta_{2n+1})|^2.\cr}}
Observe the crucial   minus sign when compared to \step. It occurs
because in one geometry the correlator depends on $\bar{z}+z'$, while
in the other on $\bar{w}-w'$.  This now converges provided  $y>y'$,
even if $x=x'=0$
ie the operators are sitting right on the boundary. Now, using the
fact
that from factors depend only on differences of rapidities,  this
expression
can be mapped with \step\ if we formally set
$\beta=\beta'+i{\pi\over 2}$, provided one has~:
\eqn\idenden{K(\beta)=R\left(i{\pi\over 2}-\beta\right),}
as claimed above.

To summarize, we can write the left right current current correlator
in two
possible ways. By using the boundary state one finds~:
\eqn\idenmaktata{
<\partial_{\bar{z}}\phi(x,y)\partial_{z'}\phi(x',y')>=
\int_0^\infty dE \ {\cal G}(E) \exp\left[E(x+x')-iE(y-y')\right],}
(recall that $x,x'<0$). One obtains ${\cal G}(E)$ simply by fixing
the energy to
a particular value in \step. When this is done, the remaining
integrations
occur on a finite domain for each of the individual particle energies
since $\sum_{i=1}^{2n+1} e^{\beta_i}=E$, and there is no problem of
convergence
anymore. One then gets~:
\eqn\nicenice{\eqalign{{\cal G}(E)
= &\sum_{n=0}^\infty \int_{-\infty}^{\ln E}
{d\beta_1\ldots
d\beta_{2n}\over(2\pi)^{2n+1}(2n+1)!}{E^2\over E-e^{\beta_1}-\ldots
-e^{\beta_{2n}}}\cr &
K(\beta_B-\beta_1)\ldots K(\beta_B-\beta_{2n})
K\left[\beta_B-\ln\left(E-e^{\beta_1}-
\ldots-e^{\beta_{2n}}\right)\right]
\cr &
\left|f\left[\beta_1\ldots\beta_{2n},\ln\left(E-e^{\beta_1}-\ldots
-e^{\beta_{2n}}
\right)\right]\right|^2,\cr}}
with the constraint $\sum_{i=1}^{2n} e^{\beta_i}\leq E$.
The denominator might suggest some possible divergences; it is
important
however to realize that it vanishes if and only if
the particle with rapidty $\beta_{2n+1}$ has vanishing energy,
in which case the form factor vanishes too. We can now shift
the integrands to write equivalently~:
\eqn\nicenicenicetata{\eqalign{{\cal  G}(E)=&E
 \sum_{n=0}^\infty \int_{-\infty}^0 {d\beta_1\ldots d\beta_{2n}
\over(2\pi)^{2n+1}(2n+1)!}{1\over 1-e^{\beta_1}-\ldots
-e^{\beta_{2n}}}\cr &
K(\ln (T_B/E)-\beta_1)\ldots K(\ln (T_B/E)-\beta_{2n})
K\left[\ln
(T_B/E)-\ln\left(1-e^{\beta_1}-\ldots-e^{\beta_{2n}}\right)\right]\cr
& \left|f\left[\beta_1\ldots\beta_{2n},
\ln\left(1-e^{\beta_1}-
\ldots-e^{\beta_{2n}}\right)\right]\right|^2,\cr}}
where the constraint
$\sum_{i=1}^{2n} e^{\beta_i}\leq 1$ is implied,  we used the fact
that form-factors depend only on rapidity differences,
and
$T_B\equiv e^{\beta_B}$.
By using the dual picture, one finds~:
\eqn\idenmaktiti{<\partial_{\bar{z}}
\phi(x,y)\partial_{z'}\phi(x',y')>=
\int_0^\infty dE {\cal F}(E) \exp\left[-iE(x+x')-E(y-y')\right],}
where~:
\eqn\nicenicenicetiti{\eqalign{{\cal  F}(E)=&-E
 \sum_{n=0}^\infty \int_{-\infty}^0 {d\beta_1\ldots d\beta_{2n}
\over(2\pi)^{2n+1}(2n+1)!}{1\over 1-e^{\beta_1}-\ldots
-e^{\beta_{2n}}}\cr &
R(\beta_1-\ln (T_B/E))\ldots R(\beta_{2n}-\ln (T_B/E))
R\left[\ln\left(1-e^{\beta_1}-\ldots-e^{\beta_{2n}}\right)-\ln
(T_B/E)
\right]\cr &
\left|f\left[\beta_1\ldots\beta_{2n},
\ln\left(1-e^{\beta_1}-
\ldots-e^{\beta_{2n}}\right)\right]\right|^2,\cr}}
where in \nicenicenicetata\ and \nicenicenicetiti\ the constraint
$\sum_{i=1}^{2n} e^{\beta_i}\leq 1$ is implied.
The two expressions are in correspondence by the simple analytic
continuation~:
\eqn\matsu{{\cal G}(E)= i{\cal F}(iE).}

\subsec{The analog of the conductance}

Although such a quantity does not have much physical
meaning, we can formally define a conductance  in the sinh-Gordon
case using the current current correlators. It is instructive to
carry out
this computation now.

To use the Kubo formula of the first section, we adopt the first
point of view
where the boundary is taken into account through the
introduction of the boundary state $|B>$. Write again~:
\eqn\idenmak{<\partial_{\bar{z}}\phi(x,y)\partial_{z'}\phi(x',0)>=
\int_0^\infty dE {\cal G}(E) \exp\left[E(x+x')-iEy\right].}
This is the only
correlation contributing to $\Delta G$ for positive Matsubara
frequencies, and
\eqn\tough{\Delta G(\omega_M)= {{\cal G}(\omega_M)\over 4\
\omega_M}.}
Here we have used the fact that $\omega_M L<<1$, ie the system,
although
large, is much
smaller than the wavelength associated with the (modulus of the) AC
frequency.

To go  to real frequencies, we can simply substitute
$\omega_M\to -i\omega$ in the $K$ matrices in the integrals
\nicenicenicetata:
\eqn\resast{\Delta G(\omega)={1\over 4\omega}\hbox {Im}{\cal
G}(-i\omega)=
{1\over 4\omega}\hbox {Re} {\cal F}(\omega).}

Recall that $K(\beta)=-\tanh(\beta/2)$.
 So we can expand the product of $K$ matrices as a series, using
\eqn\exppp{K\left[\ln(T_B/ i\omega)-\beta\right]=\left({i\omega\over
T_B}e^\beta-1\right)
 \sum_{n=0}^\infty \left(-{i\omega\over T_B}\right)^n e^{n\beta}.}
Computing term by term gives  $\Delta G$ as a power series in
$(\omega/T_B)^2$.
This
is an IR expansion, valid for   strong barriers $T_B>\omega$.

To get an UV expansion, holding for weak barriers $T_B<\omega$, we
have
to split each integration into two pieces,
$\int_{-\infty}^{-\ln(\omega/T_B)}$
and $\int_{-\ln(\omega/T_B)}^0$. In the first integral, we expand $K$
as
 in\exppp\
but in the second case we expand it as
\eqn\expppi{K(\ln(T_B/ i\omega)-\beta)=\left(1-{T_B\over
i\omega}e^{-\beta}\right)
 \sum_{n=0}^\infty \left(-{T_B\over i\omega }\right)^n e^{-n\beta}.}
This gives $\Delta G$ as a power series in $(T_B/\omega)^2$.

A nice feature of the sinh-Gordon case is that the problem is well
defined both for coupling $g$ and for its dual ${1\over g}$. This is
because
the
dimension of the perturbing operator being negative, it is always
relevant, and none of these two couplings is plagued by problems
of irrelevant perturbation theory, like what happens in the
sine-Gordon case.
We can therefore consider exactly the same problem with a coupling
$1/g$. On
the
 other
hand, these two cases are related by the sinh-Gordon duality, under
which form factors are identical, up to an overall scale (due to the
choice of $\mu$ as related with the two point function of the field
$\phi$). We
deduce therefore the identity
\eqn\ideni{G\left(g,{\omega\over T_B}\right)=g^2G\left({1\over g},
{\omega\over T_B}\right),}
where the right hand term is computed using exactly the same formula
as \step\ and \tough\  but with the formal replacement $g\to {1\over
g}$.
$T_B=e
^{\beta_B}$ is  the same in both cases. Of course, since the
correspondance between $T_B$ and the coupling $\lambda$ in \bdract\
depends
on $g$, \ideni\ maps the conductance for  $g$ and $1/g$ with
different boundary couplings. In other words, if we introduce the
function $T_B(g,\lambda)$ we have that
\eqn\idenii{G(g,\omega,\lambda)=g^2G\left({1\over
g},\omega,\lambda'\right),}
where $\lambda'$ follows from
\eqn\corrs{T_B(g,\lambda)=T_B\left({1\over g},\lambda'\right).}
Related duality properties are expected in the sine-Gordon model, but
unfortunately are much more difficult to establish.

\newsec{$\nu=1/3$ Hall effect.}

In this section we follow the same line of thought for the
sine-Gordon model.
This is the massive deformation of the free boson  which preserves
integrability with either  boundary
interactions ${\cal B}$ used in the Hall problem {\it and} the
anisotropic
Kondo model.  Thus the form factors of the sine-Gordon model in the
massless limit will be the quantities we need.
The solitons/anti-solitons and breathers quasi-excitations make the
problem
more complicated but the results presented before hold with the
addition
of a few indices (and much more complicated form factors).

\subsec{Expressions for the sine-Gordon form factors.}

We now consider the same problem in the sine-Gordon case, with
action~:
\eqn\shgoracti{S={1\over 16\pi g}\int_{-\infty}^\infty dxdy \ \left[
\left(\partial_x\phi\right)^2
+ \left(\partial_y\phi\right)^2+\Lambda\cos \phi\right].}
To compare with standard normalizations, one has $\beta^2=8\pi g$.
The form
factors approach is formally the same, albeit more complicated
because the particle content is much richer, and depends on $g$.
For $1/2 < g <1$, only solitons/anti-solitons appear in the spectrum
of the theory.  This is the so called repulsive case, with $g=1/2$
the Toulouse limit.  When $0<g<1/2$, the particle content is enriched
by $[1/g-2]$ bound states, called breathers.  In the following
we will denote by the indices $\epsilon=\pm$ the solitons and
anti-solitons, and $\epsilon=1,2,...,[1/g-2]$ the breathers.
The solitons form factors in the massive case were written by
Smirnov \Smir\
and we obtain the massless form factors by taking the appropriate
limit of the massive ones.  Only right and left moving form
factors survive in this limit, as in the sinh-Gordon case.
Moreover, the symmetry of the action dictates that only form factors
with
total topological charge zero are non-zero for the current
operator. As an example, the soliton/anti-soliton form factor
is given by~:
\eqn\ffsmir{\eqalign{
<0|\partial_z\phi(z,\bar{z})|\beta_1,\beta_2>^
{\epsilon \epsilon'}_{RR}&=\cr
\epsilon' \mu M (2\pi d)^2
e^{(\beta_1+\beta_2)/2}&{\zeta(\beta_1-\beta_2)\over
\cosh {(1-g)\over 2g}(\beta_1-\beta_2+i\pi)}\exp\left[M(e^{\beta_1}
+e^{\beta_2})z\right],}}
with $\epsilon+\epsilon'=0$
and $\epsilon=\pm$ stands for soliton (resp. antisoliton).
{}From \Smir\ one has~:
\eqn\zzzzz{\zeta(\beta)=c \sinh{\beta\over 2} \exp\left(\int_0^\infty
{\sin^2 {x(\beta+i\pi)\over 2} \sinh {\pi (1-2g)x\over 2(1-g)}\over x
\sinh {\pi g x\over 2(1-g)}\sinh\pi x\cosh {\pi x \over 2}}
dx\right),}
with the constant $c$ given by~:
\eqn\constc{
c=\left( {4(1-g)\over g}\right)^{1/4}\exp\left({1\over
4}\int_0^\infty
{\sinh {x \pi\over 2} \sinh {\pi (1-2g)x\over 2(1-g)}\over x
\sinh {\pi g x\over 2(1-g)}\cosh^2 {\pi x \over 2}} dx\right),
}
and $d$ by~:
\eqn\ddd{
d={1\over 2\pi c} {(1-g)\over g},
}
and $\mu$ is a normalization constant to be determined as before.
 Observe that in the free case $g=1/2$, the form factors \ffsmir\
reduce to
trivial kinetic terms since $\zeta(\beta)\propto \sinh(\beta/2)$.

The other soliton-antisoliton form  factors follow from the analysis
of \Smir.
Their expression simplifies in the case $g={1\over t}$, $t$ an
integer. This is the physically
relevant case for the $\nu={1\over t}$ fractional quantum Hall
effect.
One then finds~:
\eqn\ffsmirbis{\eqalign{
<0|\partial_z\phi(z,\bar{z})|\beta_1,\ldots,
\beta_{2n}>^{-,\ldots,-,+\ldots,+}_{R\ldots,R}=\mu M (8\pi^2 d)^n
e^{(\beta_1+\ldots+\beta_{2n})/2}\prod_{i<j}\zeta(\beta_i-\beta_
j)\cr
\sinh \left[{t-1\over 2} \sum_{p=1}^n
(\beta_{p+n}-\beta_p-i\pi)\right]\prod_{p=1}^n\prod_{q=n+1}^{2n}
\sinh^{-1} (t-1)
(\beta_q-\beta_p) \ det H.\cr}}
The matrix $H$ is obtained as follows. First introduce  the
function~:
\eqn\interm{\psi(\alpha)=2^{t-2}\prod_{j=1}^{t-2}\sinh
{1\over 2}\left(\alpha-i{\pi j\over t-1}+i{\pi\over 4}\right).}
One then defines the matrix elements as~:
\eqn\matrix{H_{ij}={1\over 2i\pi}\int_{-2i\pi}^0 d\alpha
\prod_{k=1}^{k=2n}
\psi(\alpha-\beta_k)\exp\left[(n-2j-1)
\alpha+(n-2i)(t-1)\alpha\right],}
where $i,j$ run over $1,\ldots, n-1$.  It is not difficult to
convince
oneself that this produce a symmetric polynomial of the right degree.
Although cumbersome, it is an easy task to extract these
determinants, as examples we find for $g=1/3$~:
\eqn\exdet{\eqalign{
det H=\exp\left(-{1\over 2} \sum_{i=1}^{2n} \beta_i\right) \
\sigma_1(e^{\beta_p}) , \ \ n=2, \cr
det H=\exp(-\sum_{i=1}^{2n} \beta_i) \ \sigma_1(e^{\beta_p})
\sigma_3(e^{\beta_p}), \ \ n=3,
\cr}}
up to irrelevant phases and
with the $\sigma_q$'s defined previously.  Having these expression
we can get all form factors using the axioms they were constructed
upon \Smir .  For example, the solitons form factors with different
positions
of the indices $\epsilon_i$ we use the symmetry property~:
\eqn\commeps{\eqalign{
f(\beta_1,...&,\beta_i,\beta_{i+1},...,\beta_n)_{\epsilon_1,...,
\epsilon_i,\epsilon_{i+1},...,\epsilon_n}
S^{\epsilon_i,\epsilon_{i+1}}_{\epsilon_i',\epsilon_{i+1}'}
(\beta_i-\beta_{i+1})\cr =&
f(\beta_1,...,\beta_{i+1},\beta_i,...,\beta_n)_{\epsilon_1,...,
\epsilon_{i+1},\epsilon_i,...,\epsilon_n} .\cr
}}
Here again, we omit the distinction between left and right moving
form
factor, they are simply related by complex conjugation.
At the points $g=1/t$ the soliton $S$ matrix used in the last
expression is reflectionless and basically just permutes the
rapidities up to a phase.  When there are breathers, the
soliton $S$ matrix has poles corresponding to the bound
states at the points $\beta=i\pi-{i\pi g\over (1-g)} m$ for the
$m$'th breather.
In view of the last relation, this induces poles
in the form factors.  We obtain the breather form factors from
these poles~:
\eqn\polerel{\eqalign{
{\rm res} f(\beta_1,...&,\beta_{n-1},\beta_n)_{\epsilon_1,...,
\epsilon_{n-1},\epsilon_n}=a_m (-1)^{{2\epsilon_n+1\over 2}m}
C_{\epsilon_{n-1},\epsilon_n} \cr
&f(\beta_1,...,\beta_{n-1}+{i\pi\over 2}-i{i\pi g\over 2 (1-g)})_{
\epsilon_1,...,\epsilon_{n-2},m}
 ,\cr
}}
and $a_m$ is given by the residue at $\beta=i\pi-{i\pi g\over (1-g)}
m$~:
\eqn\poledeux{
a_m=\left({\rm res} S_{\epsilon_{n-1}\epsilon_n}^{
\epsilon_{n-1}\epsilon_n} (\beta) \right)^{1/2}.
}
Having these relations, we posess all ingredients to compute all
form factors for $g=1/t$.  Then, using them for the computation
of the current correlations is merely an extension of the previous
results for sinh-Gordon with indices.  The normalisation of the
form factors, $\mu$,  is chosen such that \trivial\ is reproduced.
This is fixed by
introducing a complete basis of states~:
\eqn\compbas{
1=\sum_{n=0}^\infty \sum_{\epsilon_i} \int {d\beta_1 ...
d\beta_n\over
(2\pi)^n n!} |\beta_1,...,\beta_n>_{\epsilon_1,...,\epsilon_n}
{}^{\epsilon_n,...,\epsilon_1}<\beta_n,...,\beta_1|
}
and computing the correlations exactly like in the sinh-Gordon case.
It is interesting to observe however, that in the sine-Gordon case,
there
is another - a priori independent - way to fix the nromalization
$\mu$. Indeed,
the operator $\partial\phi$ being related with the $U(1)$ charge, we
need
that
\eqn\charrr{^+_R<\beta_1|\int_{-\infty}^\infty
\partial_x\phi|\beta_2>_+^R=2\pi
\delta(\beta_1-\beta_2),}
using the fact that a soliton for the bulk theory \shgoracti\ obeys
$\phi(\infty)-\phi(-\infty)=2\pi$. Using \ffsmirbis\ we get the
requirement
that
\eqn\reqi{\mu={1\over 2\pi d\zeta(-i\pi)}={2\pi g\over (1-g)}.}
Remarkably, this involves only the two particle form factor while the
requirement that \trivial\ is obeyed involves a sum over an infinity
of form-factors. However, the two should be identical if the
description is consistent,
which we checked is the case.

Knowing the normalization before performing the sum \niden\ gives us
an
a priori estimate of how many form-factors will be necessary to
compute the
full correlator. Indeed for $g=1/3$, the one breather and
2 solitons form factors normalised to $3.14$ which is very
close to the exact $\pi$.  Similarly for $g=1/4$ we found
from the contributions up to two solitons that $\mu=2.05$
to compare with $2.094=2\pi/3$.

Moreover the considerations
concerning the correlations involving the boundary state follow
in this case with now the boundary state given by~:
\eqn\bstate{
|B>=\sum_{n=0}^\infty \int_{0<\beta1<...<\beta_n}
K^{a_1b_1}(\beta_B-\beta_1)...K^{a_nb_n}(\beta_B-\beta_n)
Z_L^{*a_1}(\beta_1)Z_R^{*b_1}(\beta_1)...Z_R^{*b_n}(\beta_n),
}
with an implicit sum on the indices implied in this expression.
The matrix $K^{ab}$ is related to the boundary $R$ matrix
in the following way~:
\eqn\relkr{
K^{ab}(\beta)=R^a_{\bar{b}} \left(i{\pi\over 2}-\beta\right).
}
The $\bar{b}$ means that we take the conjugate of the indices
ie. $\pm\rightarrow \mp$ and $m\rightarrow  m$.

{}From the previous expressions, we can compute de current-current
correlation function in the presence of a boundary for $g=1/t$.
The results we will get depends on the boundary interaction, in the
next subsection we
 present some general results for all values
of $g$ when the boundary is of the form \remiv .  This is of
relevance to
the conductance in the quantum Hall effect.

\subsec{General remarks and analytic form of the conductance.}

To proceed we need the reflection matrix of the boundary
sine-Gordon theory. For generic values of the coupling $g$,
the amplitude for the processes $+\to +$ and $-\to -$ is
$R^\pm_\pm(\beta-
\beta_B)$, and for the processes $+\to -$ and $-\to +$ it is $
R^\pm_\mp(\beta-\beta_B)$
with~:
\eqn\bdrs{\eqalign{R^\pm_\mp(\beta)&={e^{(1-g)\beta\over
2g}}R(\beta)\cr
R^\pm_\pm(\beta)&= i{e^{(g-1)\beta\over 2g}}R(\beta)}}
where the function $R$ reads~:
$$\eqalign{R(\beta)&={e^{i\gamma}
\over 2\cosh\left[{(1-g) \beta\over 2g}-i{\pi\over 4}\right] }
\prod_{l=0}^\infty{Y_l(\beta)\over Y_l(-\beta)}\cr
Y_l(\beta)&={\Gamma\left({3\over 4}+l{(1-g)\over g}
-{i(1-g)\beta\over 2\pi g}\right)\Gamma\left({1\over
4}+(l+1){(1-g)\over g}
-{ i(1-g)\beta\over 2\pi g}\right)\over \Gamma\left({1\over
4}+(l+\half)
{(1-g)\over g} -{(1-g) i\beta\over 2\pi g}\right)\Gamma\left({3\over
4}+(l+\half)
{(1-g)\over g} -{ i(1-g)\beta\over 2\pi g}\right)}.\cr}$$
In \bdrs, our conventions are such that
in the UV limit ($\beta_B\to-\infty$) the scattering is totally
off-diagonal so a soliton bounces back as an anti-soliton, in
agreement with classical limit results for Neumann boundary
conditions. A useful integral representation of $R$ is given by~:
\eqn\norm{R(\beta)={e^{i\gamma}
\over 2\cosh\left[{(1-g) \beta\over 2g}-i{\pi\over 4}\right] }
\exp\left( i\int_{-\infty}^{\infty} {dy\over 2y} \sin{2(1-g) \beta
y\over g\pi}
{\sinh({1-2g\over g})y \over \sinh 2y\cosh {(1-g) y\over g}
}\right).}
Recall that the spectrum is made of one breather and the pair soliton
antisoliton in the whole domain $1/3\leq g <1/2$.
More breathers appear
for $g<1/3$, moreover the reflection matrix of the 1- breather
is the same as in the sinh-Gordon case.
There are no breathers for $g>1/2$.

In all these regimes, the form
factors are  known. They are quite complicated
for generic $g$,
and expressions for the correlators are more involved because the S
matrix
is non diagonal. We can however
extract some features of the UV and IR expansions easily, following
the same logic as in the sinh-Gordon case. To do so, consider
the soliton antisolitons reflection matrix. Evaluating the integral
in \norm\ by the residues method leads to a double expansion
of the elements $R^\epsilon_{\epsilon'}$ in powers of $\exp(\beta)$
and
$\exp({1\over g}-1)\beta$ .
This leads for the conductance to a  double power series
in $(\omega/T_B)^{-2+2/g}$  and $(\omega/T_B)^2$ in the IR,
$(T_B/\omega)^{2-2g}$  and $(T_B/\omega)^{2}$ in the UV. Breathers
do not change this result.
For instance for the 1-breather, since the reflection matrix is the
 same as in the sinh-Gordon case,
and therefore $g$ independent, the  contributions expand
as a series in $(\omega/T_B)^2$ in the IR,  $(T_B/\omega)^{2}$ in the
UV.
This holds for any coupling $g$. Therefore, as
first argued by  Guinea et al. \ref\Gui{F. Guinea, G. Gomez-Santos,
M. Sassetti, M. Ueda, Europhys. Lett. 30, 561 (1995).},  at low
frequency,
the conductance goes as $\omega^2$ for $g<1/2$, $\omega^{-2+2/g}$ for
$g>1/2$.
The $\omega^2$ power would seem
to indicate that there should be a $T^2$ term in the DC conductance,
but this
is not correct because only the modulus square of
$R^\epsilon_{\epsilon'}$
contribute to the DC conductance, and these expand only as powers
of $\exp({1\over g}-1)\beta$ .

The presence of analytic terms in the IR is a straightforward
consequence of the fact that IR perturbation theory involves
an infinity of counter-terms, in particular polynomials in
derivatives of $\phi$ \Gui. More surprising maybe is the fact
that we find analytical terms in the UV.
This requires some discussion. The UV terms follow from the short
distance behaviour of the correlation function of the current. For
any
operator
$O$ we could write formally,
\eqn\peda{<O(x',y')O(x,y)>=\sum_{n=0}^\infty (\lambda)^{2n}
\int d1\ldots dn <\tilde{O}(x',y')\tilde{O}(x,y)\cos{1\over
2}\phi(1)\ldots
\cos{1\over 2}\phi(n)>,}
where $\tilde{O}$ is the $\lambda\to 0$ limit of the field $O$. From
\peda, one would naively expect that the two point function of the
current expands as a power series in $\lambda^2$, which would lead
to a power series in $(\omega/T_B)^{2g-2}$. This is incorrect
however because, even if
 integrals are convergent at short distance for $g<1/2$, they are
always  divergent
at large distances. It is  known that these IR
divergences give precisely
 rise to non analyticity in the coupling constant $\lambda$.
One usually writes~:
\eqn\pedai{<O(x',y')O(x,y)>=\sum_i C^i_{OO}(x'-x,y'-y) O_i(x,y),}
where $O_i$ are a complete set of local operators in the theory
and the $C$'s are structure functions. These, being local quantities,
have
 analytic behaviour
in $\lambda$.  However, $<O_i(x,y)>$ being non local is in
 general non analytic - actually, on dimensional grounds,
\eqn\pedaii{<O_i(x,y)>=\lambda^{\Delta/(1-g)}\propto T_B^{\Delta},}
where $\Delta=h+\bar{h}$ is the (bulk) dimension of the field $O_i$.
If we
   computed the conductance perturbatively using Matsubara formula,
we  would
 use \peda\ with
$O$ the electrical current operator.  The case $O_i$ the identity
operator
gives rise to an analytical expression in $\lambda$, but eg the
 case $O_i=\partial_z\partial_{\bar{z}}\phi$ gives
$\lambda^{2/(1-g)}$ times
 an analytical expression in $\lambda$ (its mean value can be non
zero because there is a boundary). More generally, since the only
operators $O_i$ appearing in the case of the current
are polynomials in derivatives of $\phi$, all with integer
dimensions, we
expect that the two point function of the current will expand as a
 double series of the form $\lambda^{2n}\lambda^{2m/(1-g)}$, ie going
back
 to $T_B$ variable
, that the conductance will expand as a double series of the form
 $(T_B/\omega)^{2n(1-g)}(T_B/\omega)^{2m}$, in agreement with
the form factors result.

\subsec{The free case.}

In the case $g=1/2$ one has simply~:
\eqn\freeref{\eqalign{R^\pm_\mp(\beta)=P(\beta)&={e^\beta\over
e^\beta +i}\cr
R^\pm_\pm(\beta)=Q(\beta)&={i\over e^\beta+i}.\cr}}
In that case, only the soliton-antisoliton form factor is non zero,
$f(\beta_1,\beta_2)=\mu e^{\beta_1/2}e^{\beta_2/2}$,
where the normalization is
easily evaluated $\mu=2\pi$ and we have set $M=1$.
Hence, ${\cal F}(\omega)$ is readily evaluated
\eqn\freef{{\cal F}(\omega)=\int_{-\infty}^\infty
\int_{-\infty}^\infty
d\beta_1d\beta_2\delta(e^{\beta_1}+e^{\beta_2}-\omega)
[Q(\beta_1)Q(\beta_2)-P(\beta_1)P(\beta_2)]e^{\beta_1}e^{\beta_2},}
so
\eqn\freefi{{\cal F}(\omega)=\omega\int_0^1 dx {x(1-x)+1\over
\left(x+i{T_B\over\omega}\right)
\left(\omega-x+i{T_B\over\omega}\right)},}
from which it follows that
\eqn\freefii{\Delta G(\omega)={1\over 4}-{T_B\over
2\omega}\tan^{-1}(\omega/T_B).}
Thus we find
\eqn\freefiii{G(\omega)={1\over 2}\left(1-{T_B\over \omega}
\tan^{-1}(\omega/T_B)\right).}
This is in agreement with the solution of \KF .

\subsec{$G(\omega)$ at $g=1/3$.}

The conductance for $g=1/3$ has a direct application to the
quantum Hall effect.  Comparing with the free case, previously
treated, we now have a breather in the spectrum and non zero
form factors for all number of rapidities.  Still the convergence
is such that evaluating the first few form factors give results
to a very good accuracy, independently of the regime, UV or IR,
in which we make the computation.

In this case, the first few non zero form factors are
$f_1$, $f_{\pm,\mp}$, $f_{\pm,\mp,1}$, $f_{1,1,1}$, etc...
Here the subscript ``1" denotes the breather.  The first step is
to compute the normalisation in order to satisfy \niden .
When computing this normalisation,
we find that the first two form factors account
for the whole result to more than one percent accuracy.  Then
including
the 1 breather-2 solitons form factor is sufficient to get
the result to a very good accuracy ($F_{1,1,1}$ is
negligeable). Actually one observes that the speed of  convergence
of the form factor expansion varies  geometrically  with the number
of solitons
(counting the breathers as 2 solitons).

In order to get the
conductance we need the reflection matrices, they were given
in previous expressions and reduce to a simpler form for this value
of $g$~:
\eqn\rgtiers{
R(\beta)={1\over 2 \cosh(\beta-{i\pi\over 4})}
{\Gamma(3/8-{i\beta\over 2\pi})\Gamma(5/8+{i\beta\over 2\pi})
\over \Gamma(5/8-{i\beta\over 2\pi})\Gamma(3/8+{i\beta\over 2\pi})}
}
and the breather reflection matrix is~:
\eqn\btiers{
R^1_1(\beta)=\tanh ({\beta\over 2}-{i\pi\over 4}).}
{}From the pole of the 2 solitons form factor, the one breather
form factor is found using \polerel\ and its contributions to the
conductance
is~:
\eqn\onebr{
\Delta G(\omega)^{(1)}=-\mu^2 {\pi d^2\over 8} {\cal R}e \
\tanh({\log({\omega\over \sqrt{2} T_B})\over 2}-i\pi/4),
}
here $\mu=3.14$ is fixed by \trivial\ and $d=0.1414...$.
The contribution from the two solitons form factors is computed
similarly,
we find~:
\eqn\deuxsol{\eqalign{
\Delta G(\omega)^{(2)}=&-{\mu^2 d^2\over 2} {\cal R}e
\int_{-\infty}^0 d\beta
{R(\beta+\log({\omega\over T_B})) R(\log [(1-e^\beta){\omega\over
T_B}]\over
\cosh^2(\beta-\log(1-e^\beta))} \vert
\zeta[\beta-\log(1-e^\beta)]\vert^2
\cr &e^\beta \left[ e^\beta (1-e^\beta) ({\omega\over T_B})^2+
{1\over e^\beta (1-e^\beta) ({\omega\over T_B})^2}\right],
}}
where $\zeta(\beta)$ is the function defined in \ffsmir .
We can similarly write the following contribution, and we find that
these last
two expressions are sufficient for any reasonable purpose, they give
the
frequency dependent conductance to more than one percent accuracy.
We give the full function $G(\omega)$ in figure 5.
\fig{Frequency dependent conductance at T=0.}{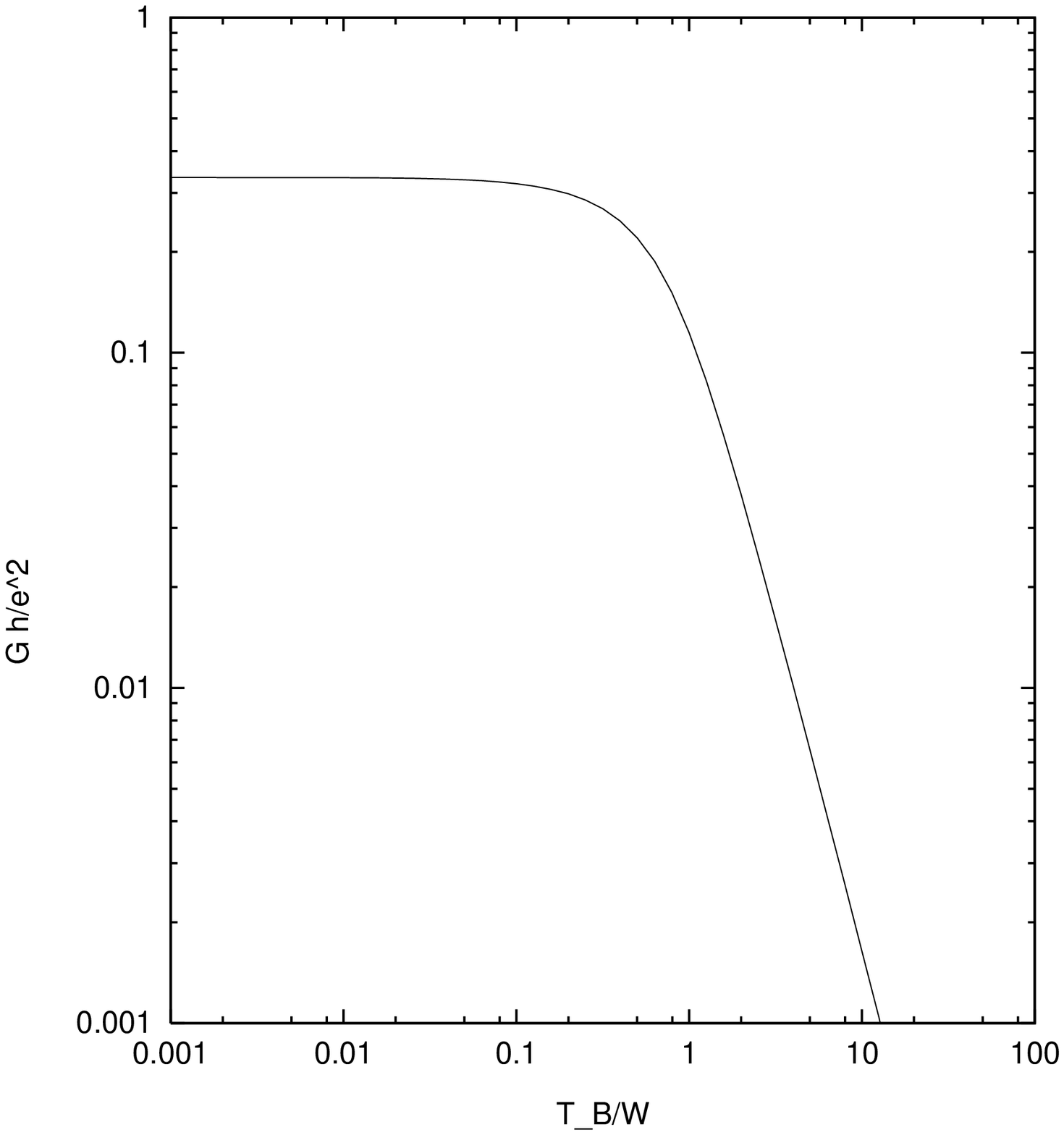}{7cm}
\figlabel\tabb

Observe that in
the UV and in the IR we obtain the $\omega$ dependance discussed
previously, even with the truncation to a few form-factors. The
form-factors
expansion is indeed  very different from the perturbative expansion
in powers of the
coupling constant in the UV, or in powers of the inverse coupling
constant
in the IR. Each form-factor contribution has by itself the same
analytical structure as the whole sum; contributions with higher
number of particles simply determine coefficients to a greater
accuracy.

\newsec{Anisotropic Kondo model and dissipative quantum mechanics.}

In section 2 we explained how the anisotropic model was related
to dissipative quantum mechanics.  The bosonised form of the
hamiltonian is~:
\eqn\hamil{H={1\over 2}\int_{-\infty}^0 dx
\left[8\pi g\Pi^2+{1\over 8\pi g}(\partial_x\phi)^2
\right]+{\lambda\over
2}\left(S_+e^{i\phi(0)/2}+
S_-e^{-i\phi(0)/2}\right),}
where $S$ are Pauli matrices.  As for the quantum Hall problem, we
keep using
as a basis the massless excitations of the sine-Gordon model; however
the boundary
interation is different.  This will result in different reflection
matrices. Another  difference with the previous section.
 will be the quantities we compute, the first step will be
to relate them to current correlation functions and then using our by
now
well known techniques to get results.

\subsec{Dissipative quantum mecanics.}

We first work in imaginary time. We consider therefore
the anisotropic Kondo problem at temperature $T$.
Let us consider the quantity
$X(y)\equiv<[S^z(y)-S^z(0)]^2>$. On the one hand, using that
 $S^z=\pm 1$,
it reads $2[1-C(y)]$, where $C(y)$ is the usual spin correlation
\eqn\corr{C(y)={1\over 2}\left[<S^z(y)S^z(0)>+
<S^z(0)S^z(y)>\right].}
On the
other hand, we can write a perturbative expansion for $X(y)$ by
expanding
evolution operators  in powers of the coupling constant $\lambda$. At
every
 order,
we get ordered monomials which are are a product of a monomial
in $S^{\pm}$ and vertex operators of charge $\pm 1/2$. We must then
evaluate $S^z(y)-S^z(0)$ for each such term, trace over the
two possible spin states, and average over the quantum field. Since
we deal
 with  spin $1/2$, terms $S^+$ and $S^-$ must
alternate, and there must be an overall equal number of
$S^+$ and $S^-$, and an equal number of  $1/2$ and $-1/2$ electric
charges.

Now, since each $S^+(y)$ comes with a $e^{-i\phi(y)/2}$
and each $S^-(y)$ comes with a $e^{i\phi(y)/2}$, $S^z(y)=S^z(0)$
if there is a vanishing electric charge inserted between $0$ and $y$,
and $S^z(y)=-S^z(0)$ if the charge inserted between $0$ and $y$
is non zero (and then it has to be $\pm 1/2$). Therefore, we can
write the
perturbation expansion of $X(y)$ in such a way that
the spin contributions all disappear:
\eqn\main{\eqalign{X(y)=&{1\over Z}\sum_{n=0}^\infty \lambda^{2n}
\sum_{alternating \epsilon_i
=\pm}
\ \sum_{p=0}^{2n} \int_0^y dy_1\int_{y_1}^y
dy_2\ldots\int_{y_{p-1}}^y dy_p
\int_y^{1/T}dy_{p+1}\ldots \int_{y_{2n-1}}^{1/T}dy_{2n}\cr
&4(\epsilon_1+\ldots+\epsilon_p)^2\left<
e^{-i\epsilon_1\phi(y_1)/2}\ldots
e^{-i\epsilon_{2n}\phi(y_{2n})/2}\right>_N,\cr}}
where $Z$ is the partition function, the factor 4 occurs
because of the normalization $S^z=\pm 1$, for every configuration of
$\epsilon$'s, only one value of $S^z(0)$ gives a non vanishing
contribution. Here, the label $N$ indicates
correlation functions for the free boson evaluated with Neumann
boundary conditions (the conditions as $\lambda\to 0$).

On the other hand,  let us consider
the correlator
\eqn\tatai{<\partial_x\phi(x,y)\phi(0,y')>_N
=-8g{x\over x^2+(y-y')^2},}
which goes to $-8g\pi\delta(y-y')$ as $x\to 0$. We have then, by
Wick's
theorem,
\eqn\tataii{\eqalign{<e^{-i\epsilon_1\phi(y_1)/2}\ldots &
e^{-i\epsilon_{2n}
\phi(y_{2n})/2}
\partial_x\phi(x,y)>_N=\cr &
8ig\left(\sum_{i=1}^{2n} \epsilon_i{x\over x^2+(y-y_i)^2}\right)
\left<e^{i\epsilon_1\phi(y_1)/2}\ldots e^
{i\epsilon_{2n}\phi(y_{2n}/2)}\right>_N,}}
and therefore
\eqn\tataiii{\eqalign{&\left<e^{-i\epsilon_1\phi(y_1)/2
}\ldots e^{-i\epsilon_{2n}\phi
(y_{2n})/2} :\partial_x\phi(x,y)
\partial_x\phi(x,y'):\right>_N=\cr
&-(8g)^2\left(\sum_{i=1}^{2n} \epsilon_i{x\over x^2+(y-y_i)^2}\right)
\left(\sum_{i=1}^{2n} \epsilon_i{x\over x^2+(y'-y_i)^2}\right)
\left<e^{-i\epsilon_1\phi(y_1)/2}\ldots
 e^{-i\epsilon_{2n}\phi(y_{2n})/2}\right>_N,\cr}}
where contractions between the dots $:$ are discarded. In \tataiii,
contractions between the dots would lead to a term factored out as
the
product of the two point function of $\partial_x\phi$  and the $2n$
point
function of vertex operators, both evaluated with N boundary
conditions. Now, we are going to
be interested in the $x\to 0$ limit where, with N boundary
conditions, $\partial_x\phi$ vanishes.   As a result we can actually
forget the subtraction in \tataiii, and write simply
obtain
\eqn\main{X(y,\lambda)=-{1\over (4g\pi)^2}\lim_{x\to 0}
\int_0^y\int_0^y dy'dy''<\partial_x\phi(x,y')\partial_x
\phi(x,y'')>_\lambda,}
where the label $\lambda$ designates the correlator
evaluated at coupling $\lambda$, N corresponding to $\lambda=0$
. Hence, we can get $C(y)$ from the current current correlator. The
latter
can then be obtained using form factors
along the above lines. The only difference is the boundary
matrix.
 If we restrict to the repulsive regime where the bulk spectrum
contains only a soliton and an antisoliton,
 one has~:
\eqn\bsol{
 R_\pm^\mp=\tanh\left({\beta\over 2}-{i\pi\over 4}\right), \ \
R^\pm_\pm=0.
}
Here again our conventions are such that
a soliton bounces back as an antisoliton, in agreement
with the    UV {\bf and}
the IR limit that have Neumann boundary conditions. In the attractive
regime we need to add the breathers with~:
\eqn\bbreath{
R^m_m={\tanh({\beta\over 2}-{i\pi m\over 4 (1/g-1)})\over
\tanh({\beta\over 2}+{i\pi m\over 4 (1/g-1)})}.
}
Writing~:
\eqn\titi{<\partial_{\bar{z}}\phi(x,y')
\partial_{z}\phi(x,y'')>_{\lambda}
=\int_0^\infty {\cal G}(E,\beta_B)\exp\left[2Ex-iE(y'-y'')\right],}
we have that~:
\eqn\titii{\eqalign{&\lim_{x\to 0}<\partial_x\phi(x,y')\partial_x
\phi(x,y'')>_\lambda=\cr
&\int_0^\infty dE\left[{\cal G}(E,\beta_B)-
{\cal G}(E,-\infty)\right]\exp[-iE(y'-y'')]+c.c. ,\cr}}
where the
 $<\partial_z\phi\partial_
z\phi>$ part and its complex conjugate (which are $\lambda$
independent) have been evaluated by requiring that
the correlator vanishes as $\lambda\to 0$ due to $N$ boundary
conditions. Hence, using the fact that ${\cal G}$ is real,
\eqn\super{X(y)={1\over 2(g\pi)^2}\int_0^\infty {dE\over
E^2}\left[{\cal G}
(E,\beta_B)-
{\cal G}(E,-\infty)\right]\sin^2 (Ey/2).}
Therefore, if we write~:
\eqn\fou{C(y)-1=\int_0^\infty A(\omega_M) \cos(\omega_M y) \
d\omega_M,}
where $\omega_M$ is a Matsubara frequency, we have~:
\eqn\mainfou{A(\omega_M)={1\over
(2g\pi)^2}{1\over\omega_M^2}\left[{\cal G}
(\omega_M,\beta_B)-
{\cal G}(\omega_M,-\infty)\right].}

An observation is now in order. From the foregoing results
we see that
\eqn\adji{<S^z(0)S^z(y)>-1={1\over (2g\pi)^2}\int_0^\infty {dE\over
E^2}
\left[{\cal G}
(E,\beta_B)-
{\cal G}(E,-\infty)\right]\cos Ey.}
On the other hand, consider the expression
\eqn\adjii{<\int_{-\infty}^0 dx'\int_{-\infty}^0 dx''
\left[<\partial_x\phi(x',y)
\partial_x\phi(x'',0)>_\lambda-<\partial_x\phi(x',y)
\partial_x\phi(x'',0)>_N\right].}
By using the same representation \titi, this is
$$
\int_{-\infty}^0 dx'\int_{-\infty}^0 dx'' \int_0^\infty dE
\left[{\cal G}
(E,\beta_B)-
{\cal G}(E,-\infty)\right] \exp\left[E(x'+x'')-iEy\right]+cc,
$$
which coincides with \adji\ after performing the integrations. We
conclude
that
\eqn\surprise{<S^z(0)S^z(y)>-1=<{\cal J}_x(0){\cal J}_x(y)>-
<{\cal J}_x(0){\cal J}_x(y)>_N,}
where we defined
\eqn\deffij{{\cal J}_x={1\over 2g\pi}\int_{-\infty}^0
\partial_x\phi(x,y) dx.}
We find also by the same manipulations that
\eqn\surprisei{<S^z(0)S^z(y)>-1=<{\cal J}_y(0){\cal J}_y(y)>-<{\cal
J}_y(0){\cal J}_y(y)>_N,}
where
\eqn\deffiij{{\cal J}_y={1\over 2g\pi}\int_{-\infty}^0
\partial_y\phi(x,y) dx.}

We now continue to real frequencies to find the response function~:
\eqn\resp{\chi''(\omega)\equiv {1\over 2}\int dt e^{i\omega t}\left<
[S^z(t),S^z(0)]\right>,}
to find~:
\eqn\mainmain{\chi''(\omega)={1\over (2g\pi)^2}{1\over\omega^2}
\hbox{Im} \left[{\cal G}
(-i\omega,\beta_B)-
{\cal G}(-i\omega,-\infty)\right].}

As a first example, let us consider the so called Toulouse limit
or  free fermion case. Then the only contribution comes from the
soliton
antisoliton
form factors, which as discussed above is $f(\beta_1,\beta_2)=\mu
e^{\beta_1/2} e^{\beta_2/2}$. Hence,
\eqn\freemain{\chi''(\omega)={1\over \pi^2} \hbox {Re}
\int_{-\infty}^\infty
\int_{-\infty}^\infty d\beta_1 d\beta_2
{e^{\beta_1}e^{\beta_2}\over (e^{\beta_1}+ie^{\beta_B})
(e^{\beta_2}+ie^{\beta_B})}
{1\over
e^{\beta_1}+e^{\beta_2}}\delta(e^{\beta_1}+e^{\beta_2}-\omega),}
that is
\eqn\freemaini{\chi''(\omega)={2\over \pi^2}{T_B\over\omega}\hbox
{Im}
 \left(
\int_0^\omega dx{1\over  (x+iT_B)(\omega-x+iT_B)}\right),}
and
\eqn\resu{\eqalign{\chi''(\omega)&={4\over \pi^2}{T_B\over\omega}
\hbox {Im}{1\over \omega+2iT_B}
 \ln\left({\omega+iT_B\over iT_B}\right)\cr
&={1\over \pi^2}
{4T_B^2\over  \omega^2 +4T_B^2}\left[{1\over\omega}
\ln\left({T_B^2+\omega^2\over T_B^2}\right)+{1\over T_B}
\tan^{-1}{\omega\over T_B}\right].\cr
}}

In general, observe that, since the reflection matrix for solitons
and antisolitons expands as a series in $e^\beta$,
${\chi''(\omega)\over\omega}$, will,
for any coupling, expand as a series of the form $(\omega/T_B)^{2n}$
in the
IR. In particular, this leads to
a behaviour $C(t)\propto {1\over t^2},t>>1$ for any $g$. In the UV,
one has to
split integrals in two pieces as explained above in \expppi. Since
the
soliton-soliton form factors expansion involves powers of $\exp
({1\over
g}-1)\beta$, ${\chi''(\omega)\over\omega}$ expands as a double series
in
$(T_B/\omega)^{2-2g}$ and $(T_B/\omega)^2$ in the UV. Hence at short
times,
$C(t)-1\propto t^{2-2g}$. This is in agreement with the qualitative
analysis of \ref\Guinea{F. Guinea, Phys. Rev. B32 (1985) 4486.}.

Results for $g\neq 1/2$ are more involved because there are non zero
form factors at all levels.  Still, when working out the first few
form
factors we observe a very rapid convergence with the number of
rapidities and again we can give precise results for different
values of $g$.  As an example, let us show the results for
$g=1/3$.

The computation for $g=1/3$ is very similar to the previous
conductance
computations.  The boundary matrices are much more simpler though.
In this case we have~:
\eqn\konsol{
R_\pm^\mp=\tanh ({\beta\over 2}-{i\pi\over 4}),\ \ R_\pm^\pm=0,
}
and~:
\eqn\konbrea{
R_1^1={\tanh({\beta\over 2}-{i\pi\over 8})\over
\tanh({\beta\over 2}+{i\pi\over 8})}.
}
Then, as was found for the conductance, we find that the first two
contributions are sufficient for most purposes, they are given
by~:
\eqn\premcontrib{
\delta\chi''(\omega)^{(1)}=-{9 \mu^2  d^2\over 8\pi  \omega }{\cal
R}e
\left[ {\tanh\left({\log({\omega\over \sqrt{2}T_B})\over
2}-{i\pi\over 8}\right)
\over
\tanh\left({\log({\omega\over \sqrt{2}T_B})\over 2}+{i\pi\over
8}\right)}
-1\right]
}
and~:
\eqn\seccontrib{\eqalign{
\delta\chi''(\omega)^{(2)}=-\left( {3 \mu d\over 2 \pi}\right)^2
{1\over \omega} {\cal R}e
\int_{-\infty}^0 &d\beta {\vert \zeta(\beta-\log
(1-e^\beta))\vert^2\over
\cosh^2(\beta-\log (1-e^\beta))} e^\beta \cr
& \left[ R^+_- (\beta+\log(\omega/T_B))
R^+_-(\log[(1-e^\beta)\omega/T_B])-1\right].\cr
}}
Again these two expressions are sufficient to get a very precise
result.
Similar computations give rise to the results in
figure 6 where we plotted  $\chi''(\omega)/\omega$  for the values,
$g=3/5,1/2, 1/3, 1/4$.
\fig{Spectral function for $T_B=0.1$.}{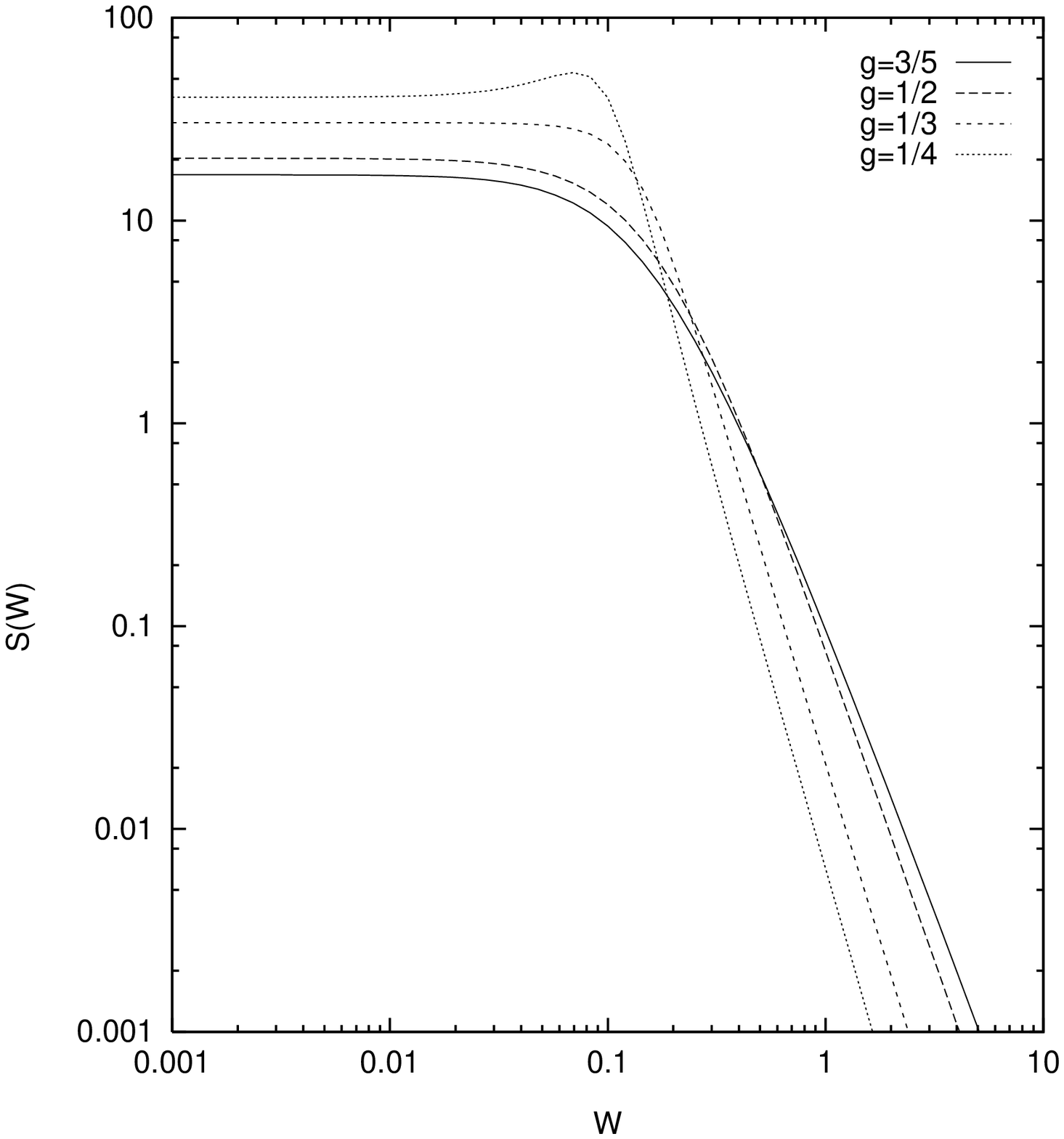}{7cm}
\figlabel\tabb

When making these calculations we have to be careful about which
terms are
needed for a good convergence.
 Our observation is that keeping the form
factors up to two rapidities give very good results precise to $1\%$.
 It is
possible to go further and get a better precision if needed.  The
last
statements are true for $g\in [0.6,0.2]$ and we believe even further
(we can get rough bounds on the higher contributions and have an
idea of the precision).  Still, for the moment the isotropic Kondo
point is difficult to treat.

A surprising result is that the emergence of quasi-particle
peaks in $S(\omega)$ is found at $g=1/3$ and not at $g=1/2$ as was
expected from other means of calculations.  Physically this
means that the behaviour of the two state system goes from
coherent to incoherent behaviour at that value of $g$.  This
is supported by a recent RG numerical study \ref\costi{
T.A. Costi, C. Kieffer, Phys. Rev. Lett. 76, (1996) 1683,
cond-mat/9601107.}.

\subsec{Shiba`s Relation.}

Up untill now, we showed results for certain values of $g$ more or
less limited by our ability (or tenacity) to write the form factors
corresponding to that value of the anisotropy, and make them
converge.  It is not impossible to find general relations though;
for  example the behaviours in the UV and the IR in different models
were infered in all generality.

Here we present a generalisation of Shiba`s relation \ref\shiba{
H. Shiba, Prog. Theor. Phys. 54, (1975) 967.} which was proven
for the Anderson model and generalised to Luttinger liquids
by Sassetti and Weiss\ref\shigen{M. Sassetti, U. Weiss,
Phys. Rev. Lett. {\bf 65}, 2262 (1990).}. The relation states that~:
\eqn\shib{
\lim_{\omega\rightarrow 0} {\chi''(\omega)\over \omega}
=2 \pi g\chi_0^2,
}
with $\chi_0$ the static succeptibility.    If we look at the
quantity~:
\eqn\cenicetata{\eqalign{{\cal  G}(E)=&E
 \sum_{n=0}^\infty \int_{-\infty}^0 {d\beta_1\ldots d\beta_{2n}
\over(2\pi)^{2n+1}(2n+1)!}{1\over 1-e^{\beta_1}-\ldots
-e^{\beta_{2n}}}\cr &
K^{a_1b_1}(\ln (T_B/E)-\beta_1)\ldots K^{a_{n-1}b_{n-1}}(\ln
(T_B/E)-\beta_{n-1})
\cr
&K^{a_nb_n}\left[\ln
(T_B/E)-\ln\left(1-e^{\beta_1}-\ldots-e^{\beta_{2n}}\right)\right]
\left|f\left[\beta_1\ldots\beta_{2n},
\ln\left(1-e^{\beta_1}-\ldots-e^{\beta_{2n}}
\right)\right]\right|^2,\cr}}
insert it in the expression for $\chi''(\omega)$
and expand it around $E\simeq 0$ we find that the contributions from
the $K$ matrices all cancel and only a constant is left (we have
to take into account the fact that the soliton/anti-solitons
$K$ matrices always appear in pair).
Then comparing
this with the UV normalisation we find that~:
\eqn\omegzero{
\lim_{\omega\rightarrow 0} {\chi''(\omega)\over \omega}=
{1\over \pi^2 g T_B^2}.
}
The total succeptibility is $\chi=\chi'+i \chi''$ and
the static succeptibility $\chi_0$ which is the
zero frequency limit of $\chi'$ can also be infered from
the previous expressions for the spin-spin correlation.
We just need to take the real part when continuing \mainfou\
to real frequencies, which leads to~:
\eqn\statsuccep{
\chi_0={1\over \pi^2 g T_B}.
}
Then in order to make contact with the previous expression
we need to renormalise the spins to $1/2$ and put the
correct normalisation\foot{We compute ${1\over 2}[S^z,S^z]$ but the
succeptibility has ${1\over 2\hbar}$ in front instead of $1/2$
thus at $h=1$ we have to renormalise by $2\pi$.}.
This amounts to multiplying the each
of the previous expressions by $\pi/2$ which leads to
the correct result~:
\eqn\shibgen{
\lim_{\omega\rightarrow 0} {\chi''(\omega)\over \omega}=
2\pi g \chi_0^2.
}

\subsec{Screening Cloud problem.}

Another problem that can be addressed using the form factor
techniques is the screening cloud problem for the anisotropic
Kondo model.  This has been a long standing problem and many
theoretical studies were devoted to it \ref\screen{
R.H. Bressmann, M. Bailyn, Phys. Rev. {\bf 154}, 471 (1967);
M.S. Fullenbaum, D.S. Falk, Phys. Rev. {\bf 157}, 452 (1967);
J. Gan, J. Phys.: Cond. Mat. {\bf 6}, 4547 (1994); K. Chen,
C. Jayaprakash, H.R. Krishnamurthy, Phys. Rev. B{\bf 45}, 5368
(1992).}.  Our one dimensional formulation here follows from
\Affleck .

Our aim here is to compute the uniform part of the
succeptibility, defined in the introduction. We restrict to the case
$g=1$
in what follows.
The electron spin density is
given by~:
\eqn\vrai{
{1\over 2}\int_{-\infty}^0 \partial_x\phi.
}
We can
proceed to the evaluation of $\chi_{un}$, which is given by~:
\eqn\chiun{
\Delta\chi_{un}(r)={1\over 8\pi}\int dt
 \ <\partial_r\phi(r)S_{tot}(t)>.
}
In the following, we will always
substract the free part ($T_B=0,\beta_B=-\infty$) and denote
the corresponding quantity by the symbol $\Delta$.
To find correct results for the zero temperature
succeptibilities, we always need to take the length of the system
to infinity first before taking $T\rightarrow 0$.  In this
case, this means that we always do the time integrals last.

Let us compute separately the static electron-electron
and electron-impurity, $\Delta\chi_{ee}$, $\Delta\chi_{ei}$
succeptibilities.
The electron-electron contribution is the appropriate integral of the
correlator
$<\partial_x\phi\partial_x\phi>-<\partial_x\phi\partial_x\phi>_N$~:
\eqn\sbos{\eqalign{
\Delta\chi_{ee}(r,T=0)&={1\over 8\pi}\int_{\infty}^\infty  dy
\int_{-\infty}^0 dx \int_0^\infty  dE \
 [{\cal G}(E,T_B)-{\cal G}(E,0)] \ e^{E(x+r)-iy}+c.c.,\cr
&={1\over 8\pi} \lim_{E\rightarrow 0}
{[{\cal G}(E,T_B)-{\cal G}(E,0)] \over E} e^{E r}, \cr
&=0, \ \  r \neq 0.
}}
with $y$ the imaginary time.  For $r=0$,
the integral over $E$ is
divergent and this computation does not make sense. In fact,
$\Delta\chi_{ee}(r,T=0)$ has a delta like contribution at $r=0$ as we
now show. Let us integrate instead $\chi_{ee}(r,T=0)$  over $r$
first, and then over $y$  since we are
at $T=0$.  The result is the static electron-electron
contribution to the succeptibility~:
\eqn\staticee{\eqalign{
\Delta\chi_{ee}(T=0)&={1\over 16\pi^2} \lim_{E\rightarrow 0}
{[{\cal G}(E,T_B)-{\cal G}(E,0)] \over E^2},\cr
&={1\over 4\pi^2 T_B}.
}}
where now the integration over $y$ made sense since the
$E$ integral was convergent. Hence we conclude
\eqn\conlcuuu{\Delta\chi_{ee}(r,T=0)={1\over 4\pi^2 T_B}\delta(r).}

Let us now come to the electron-impurity contribution and show
that it is equal and opposite to the previous contribution.
It
needs a little trick to be converted to a current correlation:
Look at the quantity~:
\eqn\trick{
{\partial_r\phi(r,y)\over 2} [S^z_{imp}(y')-S^z_{imp}(-\infty)].
}
For finite $y$, the term $<\partial_r\phi(r,y)S^z_{imp}(-\infty)>$
will
be zero because of the infinite time separation between the
operators, and the average of \trick\ will reproduce what we want.
On the
other hand, the
difference of spins can be computed  following the
transformations in section 5.1.  We then find that~:
\eqn\fithat{
{1\over 2}<\partial_r\phi(r,y)S^z_{imp}(y')>={i\over 8\pi
g}\lim_{X\rightarrow
0}
\int_{-\infty}^{y'} d\xi
<:\partial_r\phi(r,y)\partial_X\phi(X,\xi):>.
}
In order for the integral over $\xi$ to converge, we need to take
now the modular transformed representation of the current
correlation.
We finally obtain the expression~:
\eqn\spfinal{
{i\over 8\pi g}\int_0^\infty {dE\over E}
 [{\cal F}(E,T_B)-{\cal F}(E,0)] e^{-i E r+Ey'}+c.c.
}
Then using analytical continuation, we find the contribution to be
the same as the previous one  up to a sign
$\Delta\chi_{ei}(r)=-\Delta\chi_{ee}(r)$.  This is consistent with
Lowenstein \ref\lowen{
J. H. Lowenstein, Phys. Rev. B, V.29, N.7, 4120 (1984).}. This is
also
consistent with results recently found in perturbation theory
at finite $T$
\ref\barzy{V. Barzykin and I. Affleck, In progress.} showing
that these succeptibilities are zero everywhere outside $r=0$.  
This is inconsistent with the results of \barzy\ in that in their 
case, the extrapolation of the perturbative results towards zero
temperature show that {\it both} $\Delta\chi_{ie}$ and
$\Delta\chi_{ee}$ go to zero (not the sum).  This last behaviour is
also expected from a physical argument.  We don't have a clear
understanding of the discrepancy yet, since making contact between
the two calculations is not easy.   Our guess is that there is a
subtlety when the $T\rightarrow 0$ limit is taken at the same time
as the coupling to the impurity and $g\rightarrow 1$, and that 
working directly at zero temperature is not the correct way to 
proceed.
We hope to be able to verify this explicitely 
by making a finite temperature calculation in the
same formalism soon and check whether the discrepancy comes
from the limit $g\rightarrow 1$ or $T\rightarrow 0$.
In that sense, the last results, should be 
taken with caution.

All these results are related to the so called uniform part of the
succeptibility.  The $2k_f$ part, defined in section 2, involves
a different set of operators once bosonised: $e^{i\phi/2}$.
These operators have differnet anomalous dimension in the IR and UV
and
their treatment using massless form factors encounters serious
complications which are discussed in the next section.

\newsec{Form-factors failures}

For operators with naive engineering dimension such as the current,
the massless scattering approach is thus
seen to work quite nicely. Another candidate
for which things work is the  stress energy tensor, that is basically
$\left(\partial_x\phi(x)\right)^2$, describing the density of
energy
at some distance
from the boundary (the impurity). Unfortunately, things are very
different for
operators with a dimension that is not constrained by any symmetry.
An example is $\cos{1\over 2}\phi(x)$. This operator is of crucial
physical interest: for instance knowing its correlators would lead to
an exact determination
of Friedel's oscillations \ref\Fr{R. Egger, H. Grabert, Phys. Rev.
Lett. 75, 3505 (1995).}.
Unfortunately
key difficulties appear here with the form-factors approach.
Form-factors of
the operator $\cos{1\over 2}\phi$ in the bulk massive sine-Gordon
model have
been determined \Smir. A first feature is that they depend only
on rapidity differences, and do not exhibit any overall factor
depending
on the rapidity scale and fixing the (naive) dimension, as for the
current. We can then take massless limit as described above.
What one finds is
that, at least for $g={1\over integer}$,  the only form-factors
whose limit
is not cancelled exponentially by powers of $e^{-\beta_0}$ are those
which are left and right neutral - for instance form-factors
between the ground state and any state made only of breathers. A
first problem
then arises when one tries to reproduce properties of the operator
$\cos{1\over 2}\phi$ in the bulk massless theory.  For instance one
finds
that the one breather form-factor \ref\smirpap{F. Smirnov, J. Phys.
A17 (1984) L873. }is a pure, rapidity independent, number,
$$
<0|\cos{\phi\over 2}|\theta>_1=c
$$
In the massless limit, its contribution to the two point function
will therefore be of the form
$$
|c|^2\int_{-\infty}^\infty  e^{Mze^\beta}d\beta
$$
an integral that is {\bf IR divergent} - recall that for the current
it was the term giving the naive engineering dimension that made the
integrals
converge in the IR. One might think that
this problem has a simple  solution:
 put by hand the anomalous dimension, ie  multiply
in the massless limit each of the form-factors by a factor $m^{g/2}$,
where
$m$ is the bulk mass. The above contribution reads then
$$
m^g|c|^2\int_{-\infty}^\infty \exp m\left[\cosh\beta x+i\sinh\beta
y\right]d\beta
$$
Unfortunately, in the limit $m\to 0$, it does not behave any better,
and
does not reproduce the expected $|x|^{-g}$  behaviour. Recall how,
for the current, the  dimension followed  trivially from the scaling
properties of
{\bf each} individual form-factor, and was reproduced by every term.
Here,
the 1-breather example suggests that the correct behaviour can be
reproduced only when the whole series is summed up at finite $m$ and
then only $m$ is sent to zero. There is thus no hope to reproduce
even approximately the
bulk-behaviour by massless form-factors. A related situation has been
encountered in massless flows in \DMS .

We have tried to study the effect of the boundary by putting in by
hand
the correct massless bulk behaviour and concentrating on the
corrections induced by the boundary - for instance
by studying ratios of correlators with different
boundary couplings. Similar problems unfortunately occur.
 A simple
way of seing this is to consider the expected form of the
correlations of this
operator in the UV and IR limit. In the IR the field $\phi$ obeys
Dirichlet  boundary conditions, and one has
\eqn\dirprop{<\phi(z,\bar{z})\phi(z',\bar{z}'>=-4g\ln\left|{z-z'\over
\bar{z}+z'}\right|^g,}
so, restricting to arguments on the $x$ axis,
\eqn\neupropi{<\cos{\phi\over 2}(x)\cos{\phi\over
2}(x')>=\left|{x+x'\over
x-x'}\right|^g.}
In the UV on the other hand it obeys Neumann boundary conditions and
thus
\eqn\neuprop{<\cos{\phi\over 2}(x)\cos{\phi\over 2}(x')>={1\over
|x-x'|^{2g}}.}
We see thus that the anomalous dimension of the field $\cos{\phi\over
2}$ is
changing,
being equal to $g/2$ in the IR and to $g$ in the UV. This is of
course
very different from the case of derivatives of $\phi$. Thus, at the
present time, the form-factors approach fails for that operator.

\newsec{Conclusions.}

In this paper we presented a method to obtain current-current
correlations in
massless theories with interaction at the boundary.  The technique,
as
presented, is quite general and has been applied to different
problems
successfully. Several generalizations along the same lines appear
possible.
The most interesting would be to study similar  quantities but in the
presence of
a bias (or voltage) and temperature.  In that case the ground state
is
no longer empty but filled with quasi-particles in a way determined
by
the thermodynamic Bethe ansatz.  A combination of the TBA and
the technique presented here can  hopefully allow the determination
of the current-current correlation in that case.

Other generalizations appear more challenging. As explained in the
previous section, the approach, when applied to operators with a non
trivial dimension, naively fails, preventing us eg to study Friedel
oscillations for the moment. The approach also fails when  one gets
too close to the isotropic point $g=1$. Although this should not be
too catastrophic in practice because
results can be extrapolated from the $g<1$ regime, this remains  a
big challenge.

\centerline{\bf Acknowledgements.}

We thank I. Affleck and G. Mussardo for many interesting discussions.
This work
was supported by the Packard Foundation,
the National Young Investigator Program and the DOE. F. Lesage was
also  partly supported by a Canadian NSERC postdoctoral
Fellowship.

\listrefs

\bye